# Thermodynamics of amide + amine mixtures. 1. Volumetric, speed of sound and refractive index data for *N*,*N*-dimethylformamide + *N*-propylpropan-1-amine, + *N*-butylbutan-1-amine, + butan-1-amine, or + hexan-1-amine systems at several temperatures

Fernando Hevia, Ana Cobos, Juan Antonio González*, Isaías García de la Fuente, Luis Felipe Sanz

G.E.T.E.F., Departamento de Física Aplicada, Facultad de Ciencias, Universidad de Valladolid, Paseo de Belén, 7, 47011 Valladolid, Spain.

*e-mail: jagl@termo.uva.es; Fax: +34-983-423136; Tel: +34-983-423757

# Abstract

Values of density ($\rho$), speed of sound ($c$) and refractive index ($n_{\mathrm{D}}$) for *N,N*-dimethylformamide (DMF) + *N*-propylpropan-1-amine (DPA) or + butan-1-amine (BA) mixtures at (293.15-303.15) K, and for DMF + *N*-butylbutan-1-amine (DBA) or hexan-1-amine (HxA) mixtures at 298.15 K are reported. Density and speed of sound measurements were conducted using a vibrating-tube densimeter and sound analyzer, Anton Paar model DSA5000; refractive index, $n_{\mathrm{D}}$, values were obtained by means of a RFM970 refractometer from Bellingham+Stanley. The experimental $\rho$, $c$ and $n_{\mathrm{D}}$ values have been used to determine excess molar volumes, $V_{\mathrm{m}}^{\mathrm{E}}$, excess adiabatic compressibilities, $\kappa_{S}^{\mathrm{E}}$, excess speeds of sound, $c^{\mathrm{E}}$, excess thermal expansion coefficients, $\alpha_{p}^{\mathrm{E}}$, and excess refractive indices, $n_{\mathrm{D}}^{\mathrm{E}}$. This set of data show the existence of interactions between unlike molecules and of structural effects in the mixtures under study. $V_{\mathrm{m}}^{\mathrm{E}}$ values of solutions including linear secondary amines are lower than those of mixtures with linear primary amines. In fact, the contribution to $V_{\mathrm{m}}^{\mathrm{E}}$ from the breaking of amine-amine interactions is larger for the latter systems. Calculations on Rao's constant point out that there is no complex formation between the mixture components. Dispersive interactions have been analyzed by means of the molar refraction. It is shown that solutions with DPA or HxA are characterized by similar dispersive interactions and that they mainly differ in dipolar interactions.

## 1. Introduction

*N,N*-dimethylformamide (DMF) is a very polar liquid (3.7 D [1]) which is able to dissolve many organic substances, as it is an aprotic protophilic compound with excellent donor-acceptor properties. Consequently, this amide has many technical applications. For example, it is used for the production of acrylic fibers, plastics, pesticides or surface coatings [2]. In the oil industry, due to its good properties as selective extractant, it is used for the extraction of aromatic and saturated hydrocarbons and of compounds containing nitrogen [3, 4]. In addition, it results very effective in nanotechnology [5-7]. Interestingly, the detailed knowledge of liquid mixtures containing the amide functional group is essential for the understanding of complex molecules of biological interest [8]. In this context, DMF is useful as a model compound for peptides. The aqueous solution of DMF is a simple biochemical model of biological aqueous solutions [9, 10]. On the other hand, the significant local order characteristic of pure DMF and of other *N,N*-dialkylamides, related to the existence of strong dipole-dipole interactions [11], makes their theoretical study of high interest [12].

Primary and secondary amines are polar molecules (see below) which can also form hydrogen bonds giving self-associated complexes or, with the appropriate group, heterocomplexes [13-15]. Amines are also very common in Biology. In fact, the breaking of amino acids releases amines; neurotransmitters as dopamine or histamine are amines [16, 17], and the polymer DNA is usually bound to proteins which contain several amine groups [18]. In addition, many of the cations and anions of the technically important ionic liquids are related to amine groups [19].

We start this series of articles reporting density, $\rho$, data, speeds of sound, *c*, and refractive indices, $n_D$, at (293.15 K-303.15) K for DMF mixtures with *N*-propylpropan-1-amine (DPA) or butan-1-amine (BA), and at 298.15 K for DMF systems with *N*-butylbutan-1-amine (DBA) or hexan-1-amine (HxA). A literature survey shows that there are no such data for the systems under study. In contrast, volumetric [4, 20], $n_D$ [4], vapor-liquid equilibrium [21] or excess molar enthalpy [22] ($H_m^E$) measurements are available for the DMF + aniline mixture. Data on $H_m^E$ are also available for the *N*-methylethanamide + HxA system at 363.15 K [23]. The large and negative $H_m^E$ value at equimolar composition for this mixture ($-1005$ J·mol$^{-1}$) [23], and for the DMF + aniline system at 298.15 K ($-2946$ J·mol$^{-1}$) [22] reveal the existence of strong interactions between unlike molecules in amide + amine mixtures.

## 2. Experimental

*Materials.* All the compounds were used without further purification. Table 1 contains information regarding their source and purity, and Table 2 shows their physical properties, $\rho$, $c$, $n_D$, thermal expansion coefficient, $\alpha_p$, adiabatic compressibility, $\kappa_S$, and isothermal compressibility, $\kappa_T$. The values listed in Table 2 are in good agreement with the data available in the literature.

*Apparatus and procedure.* Binary mixtures were prepared by mass in small vessels of about 10 cm$^3$, using an analytical balance HR-202 (weighing accuracy 0.01 mg), with all weighings corrected for buoyancy effects. The standard uncertainty in the final mole fraction is estimated to be 0.0008. Molar quantities were calculated on the basis of the relative atomic mass table of 2015 issued by the Commission on Isotopic Abundances and Atomic Weights (IUPAC) [24].

Temperatures were measured using Pt-100 resistances, calibrated according to the ITS-90 scale of temperature, against the triple point of water and the melting point of Ga. The repeatability of the equilibrium temperature measurements is 0.01 K. The standard uncertainties for this quantity are 0.02 K and 0.03 K for $\rho$ and $n_D$ measurements, respectively (see below).

Densities and speeds of sound of both pure liquids and of the mixtures were measured by means of a vibrating-tube densimeter and sound analyzer, Anton Paar model DSA 5000, automatically thermostated within 0.01 K. A detailed description of the calibration of the apparatus has been given in an earlier work [25]. The repeatability of the $\rho$ measurements is $5 \cdot 10^{-3}$ kg $\cdot$ m$^{-3}$, while the relative standard uncertainty of the measurements is estimated to be 0.12%. The determination of the speed of sound is based on the measurement of the propagation time of short acoustic pulses (3 MHz center frequency [26]), which are repeatedly transmitted to the sample. The repeatability and standard uncertainty of the *c* measurements are, respectively, 0.1 and 0.4 m $\cdot$ s$^{-1}$. The experimental technique was checked through the determination of $V_m^E$ and $c^E$ of the (cyclohexane + benzene) mixture at (293.15-303.15) K. Our results and published values [27-29] are in good agreement. The standard uncertainty in $V_m^E$ is (0.012$|V_{m,max}^E|$ + 0.005 cm$^3$ $\cdot$ mol$^{-1}$), where $|V_{m,max}^E|$ stands for the maximum experimental value of $V_m^E$ with respect to the mole fraction. The standard uncertainty of $c^E$ is estimated to be 0.8 m $\cdot$ s$^{-1}$.

Refractive indices were measured using a refractometer model RFM970 from Bellingham+Stanley, with the temperature controlled by means of Peltier modules. The measurement technique is based on the optical detection of the critical angle at the wavelength of the sodium D line (589.6 nm). Calibration of the apparatus was undertaken using 2,2,4-trimethylpentane and toluene at (293.15-303.15) K, the working temperatures, as indicated by Marsh [30]. The temperature stability is 0.02 K, the repeatability of the $n_{\mathrm{D}}$ measurements is 0.00004 and the relative standard uncertainty is 0.0015.

## 3. Equations

The densimeter and sound analyzer Anton Paar DSA 5000 allows to obtain in straight form $\rho$, the molar volume, $V_{\mathrm{m}}$, the coefficient of thermal expansion, $\alpha_p = -(1/\rho)(\partial\rho/\partial T)_p$ and the isentropic compressibility, $\kappa_S$. As in other previous applications, $\alpha_p$ values were determined assuming that $\rho$ changes linearly with *T*. In addition, $\kappa_S$ can be determined from the Newton-Laplace equation assuming that the absorption of the acoustic wave is negligible:

$$\kappa_S = \frac{1}{\rho c^2} \tag{1}$$

The values $F^{\mathrm{id}}$ of a given thermodynamic property, *F*, for an ideal mixture at the same temperature and pressure as the investigated solution, are calculated by means of the well-established equations [31-33]:

$$F^{\mathrm{id}} = x_1 F_1^* + x_2 F_2^* \qquad (F = V_{\mathrm{m}}, C_{p\mathrm{m}}) \tag{2}$$

$$F^{\mathrm{id}} = \phi_1 F_1^* + \phi_2 F_2^* \qquad (F = \alpha_p, \kappa_T) \tag{3}$$

where $F_i^*$ is the value of the property *F* of pure component *i*, and $C_{p\mathrm{m}}$ is the molar isobaric heat capacity. In equation (3), $\phi_i = x_i V_{\mathrm{m}i}^* / V_{\mathrm{m}}^{\mathrm{id}}$ represents the volume fraction of component *i*, where $V_{\mathrm{m}i}^*$ is the molar volume of that component. Ideal values of $\kappa_S$ and *c* are calculated from the expressions [31]:

$$\kappa_S^{\mathrm{id}} = \kappa_T^{\mathrm{id}} - \frac{T V_{\mathrm{m}}^{\mathrm{id}} (\alpha_p^{\mathrm{id}})^2}{C_{p\mathrm{m}}^{\mathrm{id}}} \tag{4}$$

$$c^{\mathrm{id}} = \left( \frac{1}{\rho^{\mathrm{id}} \kappa_S^{\mathrm{id}}} \right)^{1/2} \tag{5}$$

being $\rho^{\mathrm{id}} = (x_1 M_1 + x_2 M_2)/V_{\mathrm{m}}^{\mathrm{id}}$ ($M_i$, molar mass of the *i* component). Finally, the ideal values of $n_{\mathrm{D}}$ are determined using the equation proposed by Reis *et al.* [34]:

$$n_{\mathrm{D}}^{\mathrm{id}} = \left[ \phi_1 \left(n_{\mathrm{D1}}^*\right)^2 + \phi_2 \left(n_{\mathrm{D2}}^*\right)^2 \right]^{1/2} \tag{6}$$

The excess functions are then determined from the equation:

$$F^{\mathrm{E}} = F - F^{\mathrm{id}} \qquad (F = V_{\mathrm{m}}, \kappa_S, c, \alpha_p, n_{\mathrm{D}}) \tag{7}$$

## 4. Results

Values, at the considered temperatures, $\rho$ and $c$ vs. $x_1$, the mole fraction of DMF, are collected in Table 3, while $n_{\mathrm{D}}$ results are shown in Table 4. Derived properties, as excess functions, are given in the supporting information: $V_{\mathrm{m}}^{\mathrm{E}}$ (Table S1); $\alpha_p$ and $\alpha_p^{\mathrm{E}}$ at 298.15 K (Table S2); $\kappa_S^{\mathrm{E}}$ and $c^{\mathrm{E}}$ at 298.15 K (Table S3) and $n_{\mathrm{D}}^{\mathrm{E}}$ (Table S4). These results are shown graphically in Figures 1-7. We have not found data available in the literature for comparison. The current data were fitted by unweighted least-squares polynomial regressions to the Redlich-Kister equation:

$$F^{\mathrm{E}} = x_1(1-x_1)\sum_{i=0}^{k-1} A_i (2x_1 - 1)^{\mathrm{i}} \tag{8}$$

where $F = V_{\mathrm{m}}, \kappa_S, c, \alpha_p, n_{\mathrm{D}}$. For each mixture, the number of the needed coefficients, *k*, in equation (8) was determined by applying an F-test of additional term [35] at the 99.5 % confidence level. Table 5 lists the parameters $A_i$ obtained along the adjustments, and the corresponding standard deviations $\sigma\left(F^{\mathrm{E}}\right)$, calculated from the expression:

$$\sigma\left(F^{\mathrm{E}}\right) = \left[\frac{1}{N-k}\sum\left(F_{\mathrm{cal}}^{\mathrm{E}} - F_{\mathrm{exp}}^{\mathrm{E}}\right)^2\right]^{1/2} \tag{9}$$

where *N* is the number of direct experimental values.

## 5. Discussion

Along this section, we are referring to values of the excess functions and of the thermophysical properties at 298.15 K and at $x_1$ = 0.5, except in specific cases duly indicated.

As we have previously mentioned, DMF is a very polar substance. As a consequence, its alkane mixtures show immiscibility regions up to rather high temperatures. For example, the upper critical solution temperatures of systems involving heptane or hexadecane are, respectively, 342.55 K [36] and 385.15 K [37].

Linear primary or secondary amines are weakly self-associated compounds with rather low dipole moments. For the amines considered, the values of this quantity are (in D): 1.3 (BA) [38], 1.3 (HxA) [1], 1.0 (DPA) [38], or 1.1 (DBA) [38]. $H_{\mathrm{m}}^{\mathrm{E}}$/J·mol$^{-1}$ values of mixtures including a given alkane, say heptane, are: 1192 (BA) [39], 962 (HxA) [39], 424 (DPA) [40], and 317 (DBA) [40]. These positive $H_{\mathrm{m}}^{\mathrm{E}}$ values can be explained in terms of the disruption of amine-amine interactions along the mixing process. We note that $H_{\mathrm{m}}^{\mathrm{E}}$ decreases when the self-association of the amine becomes weaker, as the amine group is more sterically hindered in longer amines, and in secondary amines than in primary amines. On the other hand, it is well stated that positive $V_{\mathrm{m}}^{\mathrm{E}}$ values are related to the breaking of interactions between like molecules, while negative values come from the creation of solute-solvent interactions and/or structural effects (geometrical factors including differences in size and shape between the mixture compounds [41-43] or interstitial accommodation [44]). The $V_{\mathrm{m}}^{\mathrm{E}}$ (heptane)/cm$^{3}$·mol$^{-1}$ values are: 0.7171 (BA) [45], 0.3450 (HxA) [45], 0.2752 (DPA) [46], and 0.0675 (DBA) [46]. Interestingly, the $H_{\mathrm{m}}^{\mathrm{E}}$ and $V_{\mathrm{m}}^{\mathrm{E}}$ values are positive and change in line, which reveals that the most important contribution to $V_{\mathrm{m}}^{\mathrm{E}}$ comes from the disruption of amine-amine interactions upon mixing. However, structural effects may also be present. The low $V_{\mathrm{m}}^{\mathrm{E}}$ value of DBA + heptane system, and the negative value of the DBA + hexane mixture (–0.185 cm$^{3}$·mol$^{-1}$) [47] support this statement, as positive $H_{\mathrm{m}}^{\mathrm{E}}$ values and those negative of $V_{\mathrm{m}}^{\mathrm{E}}$ for a given solution suggest that the most relevant contribution to the latter excess function arises from structural effects [43].

In view of the mentioned features, the negative $V_{\mathrm{m}}^{\mathrm{E}}$ /cm$^{3}$·mol$^{-1}$ values of DMF + amine mixtures for systems containing DPA (–0.289), BA (–0.263) or HxA (–0.021) and the low $V_{\mathrm{m}}^{\mathrm{E}}$ positive value for the DBA solution (0.018 cm$^{3}$·mol$^{-1}$) can be ascribed to the existence of DMF-amine interactions as well as to structural effects. It must be noted that the increase of the amine size along a homologous series leads to increased $V_{\mathrm{m}}^{\mathrm{E}}$ values. This means that the contributions that increase $V_{\mathrm{m}}^{\mathrm{E}}$ (larger number of DMF-DMF interactions broken by the longer amines and the weakening of the amide-amine interactions related to the fact that the amine group is more sterically hindered in such amines) are predominant over those decreasing $V_{\mathrm{m}}^{\mathrm{E}}$ (difference in size between components, lower positive contribution from the disruption of the amine-amine interactions).

The replacement of a primary linear amine (HxA) by a linear secondary amine (DPA) leads to decreased $V_{\mathrm{m}}^{\mathrm{E}}$ values. It is remarkable that the same behavior is encountered for HxA or DPA + heptane mixtures (see above). Therefore, the observed variation in DMF solutions can be ascribed to a lower positive contribution to $V_{\mathrm{m}}^{\mathrm{E}}$ from the breaking of the amine-amine interactions. A similar trend is encountered in 1-alkanol + HxA, or + DPA systems [48, 49]. The more negative $V_{\mathrm{m}}^{\mathrm{E}}$ value of the DMF + aniline mixture (–0.6931 $\mathrm{cm^3{\cdot}mol^{-1}}$) [20] compared to those of the systems with HxA or DPA suggests that the presence of an aromatic ring leads to stronger interactions between unlike molecules, which is in agreement with the largely negative $H_{\mathrm{m}}^{\mathrm{E}}$ value of this system (see Introduction). Solutions including DPA or BA show negative values of $A_p = \left(\Delta V_{\mathrm{m}}^{\mathrm{E}} / \Delta T\right)_p$ and $\alpha_p^{\mathrm{E}}$ (Table S2). Thus, the use of $V_{\mathrm{m}}^{\mathrm{E}}$ ($x_1$ = 0.5) values obtained at different temperatures gives $A_p$ /$\mathrm{cm^3{\cdot}mol^{-1}{\cdot}K^{-1}}$ = $-1.8{\cdot}10^{-3}$ (DPA); $-2{\cdot}10^{-4}$ (BA). This means that the structure of the mixture is more difficult to be broken than that of the pure liquids, which may be considered as an evidence of the existence of interactions between unlike molecules. In fact, values of $A_p$ and $\alpha_p^{\mathrm{E}}$ are positive at any composition for solutions where strong interactions between like molecules are present. This is the case, e.g, of the 2-ethoxyethanol + octane [50] or the pentan-1-ol + cyclohexane [51] systems ($A_p$ /$\mathrm{cm^3{\cdot}mol^{-1}{\cdot}K^{-1}}$ = $7.6{\cdot}10^{-3}$; $2.3{\cdot}10^{-3}$, respectively). However, $A_p$ values are also negative for solutions characterized by relevant structural effects ($-1.3{\cdot}10^{-2}$ $\mathrm{cm^3{\cdot}mol^{-1}{\cdot}K^{-1}}$ for the hexane + hexadecane mixture [52]). Taking into account the different molar volumes of DPA (137.93 $\mathrm{cm^3{\cdot}mol^{-1}}$) and BA (99.88 $\mathrm{cm^3{\cdot}mol^{-1}}$), the more negative $A_p$ value of the DPA system may be related, at least partially, to structural effects. On the other hand, $A_p$ (DMF + aniline) [20] = $-2.9{\cdot}10^{-3}$ $\mathrm{cm^3{\cdot}mol^{-1}{\cdot}K^{-1}}$, which is a more negative value than that of the DPA solution. This supports our previous statement, that DMF-amine interactions are stronger in the aniline system. The $\kappa_S^{\mathrm{E}}$ values can be also interpreted in terms of structural and interactional effects [53]. Structural effects and interactions between unlike molecules lead to negative values of this magnitude ($\kappa_S^{\mathrm{E}}$ /$\mathrm{TPa^{-1}}$= $-142$ (aniline + propanone) [54]). Positive values are encountered in solutions where interactions between like molecules are predominant ($\kappa_S^{\mathrm{E}}$ /$\mathrm{TPa^{-1}}$ = 15.3 (2-ethoxyethanol + *n*-octane) [55]). For the systems under study, $\kappa_S^{\mathrm{E}}$ /$\mathrm{TPa^{-1}}$ = $-47.8$ (DPA); $-17.7$ (DBA); $-41.9$ (BA); $-13.3$ (HxA), which is consistent with the trends mentioned above. In addition, the consistency between the signs of the $V_{\mathrm{m}}^{\mathrm{E}}$, $\kappa_S^{\mathrm{E}}$ and $c^{\mathrm{E}}$ functions must be remarked, as $V_{\mathrm{m}}^{\mathrm{E}}$, $\kappa_{\mathrm{S}}^{\mathrm{E}}$ are negative and $c^{\mathrm{E}}$ is positive (Tables S1 and S3; Figures 1-6). The DBA mixture slightly separates from this trend and $V_{\mathrm{m}}^{\mathrm{E}}$ is small and positive. However, we underline the strong asymmetry of the $\kappa_S^{\mathrm{E}}$ curve, with a minimum in the region where $V_{\mathrm{m}}^{\mathrm{E}}$ shows negative values (Figures 1 and 3).

We have also determined the internal pressures, $P_{\mathrm{int}}$ [56-59]:

$$P_{\mathrm{int}} = \frac{\alpha_p T}{\kappa_T} - p \tag{10}$$

and the excess internal pressures, $P_{\mathrm{int}}^{\mathrm{E}} = P_{\mathrm{int}} - P_{\mathrm{int}}^{\mathrm{id}}$, with $P_{\mathrm{int}}^{\mathrm{id}} = \alpha_p^{\mathrm{id}} T \big/ \kappa_T^{\mathrm{id}} - p$ [60]. The $\kappa_T$ values of the mixtures were obtained from

$$\kappa_T = \kappa_S + \frac{T V_{\mathrm{m}} \alpha_p^2}{C_{p,\mathrm{m}}} \tag{11}$$

assuming that $C_{p\mathrm{m}}^{\mathrm{E}} = 0$, and that $\alpha_p = \alpha_p^{\mathrm{id}}$ (equation 3) when experimental data are not available. For pure compounds, we have $P_{\mathrm{int}}$/MPa = 455.7 (DMF); 303.9 (DPA); 306.7 (DBA); 339.2 (BA); 345 (HxA), and for the DMF mixtures, $P_{\mathrm{int}}$/MPa = 353.9 (DPA); 345.9 (DBA); 389.4 (BA); 382.2 (HxA). Because the main contributions to $P_{\mathrm{int}}$ are related to dispersion forces and weak dipole-dipole interactions [58], these values suggest that dipolar interactions between unlike molecules are more relevant in systems including linear primary amines. On the other hand, $P_{\mathrm{int}}^{\mathrm{E}}$/MPa = 14.6 (DPA); 6.5 (DBA); 14.4 (BA); 5.9 (HxA). Large positive $P_{\mathrm{int}}^{\mathrm{E}}$ values are encountered in systems characterized by strong interactions between unlike molecules. For example, $P_{\mathrm{int}}^{\mathrm{E}}$(aniline + propanone) = 61.4 MPa [54]. It is rather clear that the higher $P_{\mathrm{int}}^{\mathrm{E}}$ value of the DPA system compared to that of the HxA mixture cannot be ascribed to stronger interactions between unlike molecules but to structural effects.

On the other hand, $P_{\mathrm{int}}$ values can be calculated using the equation [57]:

$$P_{\mathrm{int}} = \frac{RT}{x_1 v_{\mathrm{f}1} + x_2 v_{\mathrm{f}2} + V_{\mathrm{m}}^{\mathrm{E}}} - p \tag{12}$$

In this expression, $v_{\mathrm{f}i}$ denotes the molar free volume of component *i*, obtained from $v_{\mathrm{f}i} = RT/(p + P_{\mathrm{int},i})$ [57]. Results on $P_{\mathrm{int}}$/MPa from equation (12) are: 380.7 (DPA), 365.7 (DBA), 389.4 (BA) and 394.1 (HxA). The differences with the experimental values (equation. 10) are: 7.6%, 5.7%, 4.2% and 3.1%, respectively. This demonstrates that the van der Waals equation holds to a rather large extent for the investigated solutions, as equation (12) is derived from this equation of state [57].

The Rao's constant [61], $R_c$ , (also termed molar sound velocity, $R_c = V_m c^{1/3}$) is a quantity commonly used to investigate molecular interactions in liquid mixtures from ultrasonic measurements. In fact, if there is no association, or if the degree of association does not depend on concentration, $R_c$ changes linearly on the mole fractions of the components and one can write [62-64]: $R_c = x_1 R_{c1} + x_2 R_{c2}$ Systems where complex formation is present show deviations from this behavior [64]. For the actual mixtures under study, $R_c$ varies linearly with $x_1$ (Figure 8), and this indicates that there is no complex formation [62, 63].

Finally, the $n_D$ values can be used for the determination of the molar refraction $R_m$, a quantity closely related to the dispersion forces of the considered system, as $n_D$ at optical wavelengths is related to the mean electronic polarizability [65]. $R_m$ can be calculated using the Lorentz-Lorenz equation [65, 66]:

$$R_m = \frac{n_D^2 - 1}{n_D^2 + 2} V_m \tag{13}$$

We have $R_m$ (DMF)/cm$^3$·mol$^{-1}$ = 26.7 (DPA); 31.4 (DBA); 22.0 (BA); 26.7 (HxA). These results allow to state that: (i) as expected, dispersive interactions become more relevant when the amine size increases along a homologous series; (ii) dispersive interactions are more or less similar in DPA and HxA mixtures, which means that such solutions mainly differ in dipolar interactions.

## 6. Conclusions

Data on $\rho$, $c$ and $n_D$ for DMF + DPA, + DBA, + BA or + HxA mixtures at different temperatures have been reported, and the excess functions $V_m^E$ , $\kappa_S^E$ , $c^E$ , $\alpha_p^E$ and $n_D^E$ have been calculated. The data show the existence of interactions between unlike molecules and of structural effects in the investigated systems. $V_m^E$ values of mixtures including linear secondary amines are lower than those of systems with linear primary amines, as for the latter solutions the contribution to $V_m^E$ from the breaking of amine-amine interactions is larger. Mixtures with DPA or HxA differ essentially in dipolar interactions.

## Acknowledgements

F. Hevia acknowledges the grant received from the program 'Ayudas para la Formación de Profesorado Universitario (convocatoria 2014), de los subprogramas de Formación y de Movilidad incluidos en el Programa Estatal de Promoción del Talento y su Empleabilidad, en el marco del Plan Estatal de Investigación Científica y Técnica y de Innovación 2013-2016, de la Secretaría de Estado de Educación, Formación Profesional y Universidades, Ministerio de Educación, Cultura y Deporte, Gobierno de España'.

## Supporting information

This material contains values of $V_{\mathrm{m}}^{\mathrm{E}}$ and $n_{\mathrm{D}}^{\mathrm{E}}$ at the working temperatures and values of $\alpha_p$, $\alpha_p^{\mathrm{E}}$, $\kappa_S^{\mathrm{E}}$, $c^{\mathrm{E}}$ at 298.15 K.

Table 1

Sample description.

| Chemical name | CAS number | Source | Purification method | Mole fraction purity | Analysis method |
|---|---|---|---|---|---|
| *N*,*N*-dimethylformamide (DMF) | 68-12-2 | Fluka | none | ≥ 0.995 | GC[a] |
| *N*-propylpropan-1-amine (DPA) | 142-84-7 | Fluka | none | ≥ 0.99 | GC[a] |
| *N*-butylbutan-1-amine (DBA) | 111-92-2 | Aldrich | none | ≥ 0.995 | GC[a] |
| butan-1-amine (BA) | 109-73-9 | Sigma-Aldrich | none | ≥ 0.99 | GC[a] |
| hexan-1-amine (HxA) | 111-26-2 | Aldrich | none | ≥ 0.995 | GC[a] |

[a]Gas-liquid chromatography.

Table 2

Physical properties of pure compounds at temperature $T$ and pressure $p$ = 0.1 MPa. [a]

| Property | $T$/K | DMF | DPA | DBA | BA | HxA |
|---|---|---|---|---|---|---|
| $\rho^*$ /g·cm$^{-3}$ | 293.15 | 0.948881<br>0.948922[b] | 0.738194<br>0.738188[c] | 0.759695<br>0.759571[c] | 0.737048 | 0.764423 |
| | 298.15 | 0.944081<br>0.944163[b] | 0.733618<br>0.733683[c] | 0.755525<br>0.755457[c] | 0.732231<br>0.7327[d] | 0.760073<br>0.76013[e] |
| | 303.15 | 0.939361<br>0.939390[b] | 0.729098<br>0.729087[c] | 0.751458<br>0.751329[c] | 0.727452 | 0.755848 |
| $c^*$/m·s$^{-1}$ | 293.15 | 1476.8<br>1477.8[b] | 1209.4<br>1209[c] | 1261.1<br>1261.2[c] | 1268.3 | 1324.0 |
| | 298.15 | 1457.2<br>1458.5[b]<br>1458.6[g] | 1187.7<br>1198[f] | 1241.5<br>1248[f] | 1246.0<br>1247.8[d] | 1303.6<br>1304.7[e] |
| | 303.15 | 1438.2<br>1439[b]<br>1440.3[g] | 1167.2<br>1174[f] | 1222.5<br>1227[f] | 1224.6<br>1227[f] | 1283.6<br>1285[f] |
| $\alpha_p^*$ /10$^{-3}$K$^{-1}$ | 298.15 | 1.008<br>1.010[g] | 1.240<br>1.29[h] | 1.090<br>1.12[h] | 1.311<br>1.314[f] | 1.128<br>1.13[e] |
| $\kappa_S^*$ /TPa$^{-1}$ | 293.15 | 483.2<br>485[b] | 926.2<br>926.5[f] | 827.7 | 843.4 | 746.3 |
| | 298.15 | 498.8<br>498.7[i]<br>497.9[b] | 966.3<br>947[f] | 858.7<br>849[f] | 879.7<br>876.6[d] | 774.2<br>773[e] |
| | 303.15 | 514.7<br>514[b]<br>512.9[g] | 1006.7<br>992[f] | 890.4<br>883[f] | 916.7<br>912[f] | 802.9<br>800[f] |
| $\kappa_T^*$ /TPa$^{-1}$ | 298.15 | 659.4<br>650[h]<br>662[j] | 1216.4<br>1183[f] | 1059.4<br>1039[f] | 1151.9<br>1145[f] | 974.6<br>975[e] |
| $C_{pm}^*$ /J·mol$^{-1}$·K$^{-1}$ | 298.15 | 146.05[k] | 252.84[h] | 302[f] | 188[l] | 252[l] |
| $n_D^*$ | 293.15 | 1.43055<br>1.43047[h]<br>1.4281[m] | 1.40432<br>1.4043[h] | 1.41724<br>1.4177[h] | 1.40060<br>1.40106[n] | |
| | 298.15 | 1.42828<br>1.42817[h]<br>1.4280[j] | 1.40139<br>1.4053[f] | 1.41488<br>1.4152[h] | 1.39786<br>1.3987[h] | 1.41577<br>1.4160[f] |
| | 303.15 | 1.42603<br>1.4267[o]<br>1.4271[j] | 1.39883<br>1.4022[f] | 1.41253<br>1.4143[f] | 1.39500<br>1.3978[f]<br>1.39744[n] | |

[a] $\rho^*$, density; $c^*$, speed of sound; $\alpha_p^*$, isobaric thermal expansion coefficient; $\kappa_S^*$, adiabatic compressibility; $\kappa_T^*$, isothermal compressibility; $C_{p\mathrm{m}}^*$, isobaric molar heat capacity; and $n_\mathrm{D}^*$, refractive index. Standard uncertainties, $u$, are: $u(T) = 0.02\ \mathrm{K}$ (for $n_D^*$ values, $u(T) = 0.03\,\mathrm{K}$); $u(p) = 1\ \mathrm{kPa}$; $u(c^*) = 0.4\ \mathrm{m \cdot s^{-1}}$. Relative standard uncertainties, $u_\mathrm{r}$, are: $u_\mathrm{r}(\rho^*) = 0.0012$; $u_\mathrm{r}(\alpha_p^*) = 0.028$; $u_\mathrm{r}(\kappa_S^*) = 0.002$; $u_\mathrm{r}(\kappa_T^*) = 0.015$; $u_\mathrm{r}(n_D^*) = 0.0015$. [b]Ref. [67]; [c]Ref. [68]; [d]Ref. [69]; [e]Ref. [70]; [f]Ref. [71]; [g]Ref. [72]; [h]Ref. [73]; [i]Ref. [74]; [j]Ref. [75]; [k]Ref. [76]; [l]Ref. [77]; [m]Ref. [78]; [n]Ref. [79]; [o]Ref. [80].

Table 3

Densities, $\rho$, and speeds of sound, $c$, for *N*,*N*-dimethylformamide (1) + amine (2) mixtures at temperature $T$ and pressure $p$ = 0.1 MPa. [a]

| $x_1$ | $\rho$ /g·cm$^{-3}$ | $c$ /m·s$^{-1}$ | $x_1$ | $\rho$ /g·cm$^{-3}$ | $c$ /m·s$^{-1}$ |
|---|---|---|---|---|---|
| DMF (1) + DPA (2) ; $T$/K = 293.15 K | | | | | |
| 0.0600 | 0.745894 | 1218.2 | 0.4974 | 0.815656 | 1299.2 |
| 0.1071 | 0.752141 | 1225.0 | 0.5487 | 0.825989 | 1312.0 |
| 0.1560 | 0.758933 | 1232.9 | 0.6562 | 0.849651 | 1342.1 |
| 0.1975 | 0.764870 | 1239.3 | 0.7514 | 0.873097 | 1373.4 |
| 0.2490 | 0.772540 | 1248.1 | 0.8216 | 0.892212 | 1399.8 |
| 0.3099 | 0.782194 | 1259.4 | 0.8501 | 0.900523 | 1411.2 |
| 0.3463 | 0.788151 | 1266.2 | 0.9025 | 0.916508 | 1433.3 |
| 0.3985 | 0.797199 | 1276.9 | 0.9486 | 0.931283 | 1453.4 |
| 0.4480 | 0.806192 | 1287.6 | | | |
| DMF (1) + DPA (2) ; $T$/K = 298.15 K | | | | | |
| 0.0626 | 0.741577 | 1196.9 | 0.5386 | 0.819199 | 1288.9 |
| 0.1083 | 0.747697 | 1204.0 | 0.6082 | 0.833997 | 1307.9 |
| 0.1544 | 0.754025 | 1211.2 | 0.6527 | 0.844084 | 1321.1 |
| 0.2541 | 0.768732 | 1228.1 | 0.7477 | 0.867448 | 1352.4 |
| 0.3148 | 0.778293 | 1239.3 | 0.8063 | 0.883226 | 1374.0 |
| 0.3609 | 0.786012 | 1248.3 | 0.9020 | 0.911457 | 1413.1 |
| 0.4077 | 0.794223 | 1258.2 | 0.9482 | 0.926367 | 1433.5 |
| 0.4966 | 0.810800 | 1278.5 | | | |
| DMF (1) + DPA (2) ; $T$/K = 303.15 K | | | | | |
| 0.0453 | 0.734885 | 1174.0 | 0.5415 | 0.815193 | 1270.3 |
| 0.1034 | 0.742550 | 1183.0 | 0.6016 | 0.827909 | 1286.6 |
| 0.1963 | 0.755529 | 1198.0 | 0.6517 | 0.839209 | 1301.4 |
| 0.2582 | 0.764775 | 1208.8 | 0.7502 | 0.863388 | 1334.0 |
| 0.3559 | 0.780614 | 1227.6 | 0.8498 | 0.890941 | 1372.0 |
| 0.4099 | 0.790030 | 1239.0 | 0.9000 | 0.906165 | 1393.2 |
| 0.4565 | 0.798539 | 1249.4 | 0.9498 | 0.922203 | 1415.2 |
| DMF (1) + DBA (2) ; $T$/K = 298.15 K | | | | | |
| 0.0642 | 0.761171 | 1246.4 | 0.5574 | 0.823977 | 1307.9 |
| 0.1134 | 0.765782 | 1250.5 | 0.5996 | 0.831688 | 1316.3 |
| 0.1647 | 0.770882 | 1255.0 | 0.6514 | 0.842013 | 1328.1 |
| 0.2121 | 0.775872 | 1259.5 | 0.6869 | 0.849571 | 1336.8 |
| 0.2734 | 0.782835 | 1266.1 | 0.7361 | 0.860924 | 1350.4 |
| 0.3213 | 0.788624 | 1271.6 | 0.7904 | 0.874714 | 1367.5 |
| 0.4114 | 0.800702 | 1283.5 | 0.8418 | 0.889103 | 1385.9 |
| 0.4526 | 0.806766 | 1289.7 | 0.8921 | 0.904666 | 1406.1 |
| 0.5072 | 0.815344 | 1298.6 | 0.9471 | 0.923650 | 1431.0 |
| DMF (1) + BA (2) ; *T*/K = 293.15 K | | | | | |
| 0.0599 | 0.747390 | 1277.7 | 0.5483 | 0.842154 | 1368.9 |
| 0.1056 | 0.755445 | 1285.0 | 0.6569 | 0.866102 | 1393.7 |
| 0.1591 | 0.765035 | 1293.8 | 0.6999 | 0.875844 | 1404.0 |
| 0.2505 | 0.782005 | 1309.6 | 0.7537 | 0.888314 | 1416.7 |
| 0.3017 | 0.791811 | 1318.9 | 0.8036 | 0.900101 | 1429.0 |
| 0.3583 | 0.802880 | 1329.6 | 0.8579 | 0.913315 | 1442.3 |
| 0.3992 | 0.811114 | 1337.8 | 0.9048 | 0.924751 | 1453.7 |
| 0.5059 | 0.833130 | 1359.6 | 0.9546 | 0.937212 | 1465.6 |
| DMF (1) + BA (2) ; *T*/K = 298.15 K | | | | | |
| 0.0575 | 0.742224 | 1255.4 | 0.4381 | 0.814268 | 1324.9 |
| 0.1092 | 0.751333 | 1263.9 | 0.5028 | 0.827736 | 1338.6 |
| 0.1519 | 0.759015 | 1271.1 | 0.6005 | 0.848795 | 1360.6 |
| 0.2043 | 0.768674 | 1280.1 | 0.6934 | 0.869781 | 1382.5 |
| 0.2448 | 0.776215 | 1287.4 | 0.7572 | 0.884516 | 1398.0 |
| 0.3090 | 0.788510 | 1299.3 | 0.8056 | 0.895958 | 1410.0 |
| 0.3558 | 0.797705 | 1308.4 | 0.9076 | 0.920837 | 1435.1 |
| 0.3986 | 0.806285 | 1317.0 | 0.9553 | 0.932723 | 1446.6 |
| DMF (1) + BA (2) ; *T*/K = 303.15 K | | | | | |
| 0.0505 | 0.736213 | 1233.0 | 0.5281 | 0.828281 | 1324.0 |
| 0.1477 | 0.753539 | 1249.4 | 0.6056 | 0.845104 | 1341.6 |
| 0.2019 | 0.763466 | 1259.0 | 0.6960 | 0.865473 | 1363.2 |
| 0.2410 | 0.770708 | 1266.0 | 0.7558 | 0.879286 | 1377.7 |
| 0.3024 | 0.782423 | 1277.5 | 0.8006 | 0.889962 | 1389.0 |
| 0.3558 | 0.792882 | 1287.8 | 0.8544 | 0.902880 | 1402.3 |
| 0.4358 | 0.808960 | 1304.1 | 0.8998 | 0.914033 | 1413.7 |
| 0.5114 | 0.824673 | 1320.3 | 0.9466 | 0.925732 | 1425.2 |
| DMF (1) + HxA (2) ; *T*/K = 298.15 K | | | | | |
| 0.0506 | 0.765582 | 1307.1 | 0.6022 | 0.846551 | 1368.3 |
| 0.0992 | 0.771096 | 1310.8 | 0.7056 | 0.867688 | 1387.1 |
| 0.1725 | 0.779897 | 1316.7 | 0.7996 | 0.889217 | 1406.8 |
| 0.2548 | 0.790546 | 1324.2 | 0.8492 | 0.901602 | 1418.4 |
| 0.3476 | 0.803611 | 1333.7 | 0.8982 | 0.914549 | 1430.4 |
| 0.4461 | 0.818861 | 1345.5 | 0.9530 | 0.929990 | 1444.7 |
| 0.5500 | 0.836751 | 1360.1 | | | |

[a] The standard uncertainties, $u$, are: $u(x_1) = 0.0008$; $u(p) = 1$ kPa; $u(T) = 0.02$ K. The combined expanded standard uncertainties (0.95 level of confidence) are: $U_{rc}(\rho) = 0.0024$ (relative value); $U_c(c) = 0.8\,\mathrm{m \cdot s^{-1}}$.

Table 4

Refractive indices, $n_D$, of *N,N*-dimethylformamide (1) + amine (2) mixtures at temperature $T$ and pressure $p$ = 0.1 MPa. [a]

| $x_1$ | $n_D$ | $x_1$ | $n_D$ |
|---|---|---|---|
| DMF (1) + DPA (2) ; $T$/K = 293.15 K | | | |
| 0.0600 | 1.40543 | 0.5487 | 1.41645 |
| 0.1556 | 1.40725 | 0.6018 | 1.41789 |
| 0.2614 | 1.40952 | 0.7033 | 1.42079 |
| 0.3463 | 1.41140 | 0.8216 | 1.42444 |
| 0.3985 | 1.41266 | 0.8844 | 1.42653 |
| 0.4554 | 1.41402 | 0.9486 | 1.42882 |
| DMF (1) + DPA (2) ; $T$/K = 298.15 K | | | |
| 0.0626 | 1.40259 | 0.4966 | 1.41247 |
| 0.1083 | 1.40358 | 0.6527 | 1.41673 |
| 0.1544 | 1.40443 | 0.7477 | 1.41960 |
| 0.2541 | 1.40662 | 0.8063 | 1.42147 |
| 0.3148 | 1.40796 | 0.9020 | 1.42478 |
| 0.3609 | 1.40908 | 0.9482 | 1.42645 |
| 0.4077 | 1.41017 | | |
| DMF (1) + DPA (2) ; $T$/K = 303.15 K | | | |
| 0.0453 | 1.39971 | 0.6016 | 1.41295 |
| 0.1034 | 1.40091 | 0.7502 | 1.41745 |
| 0.1963 | 1.40293 | 0.8498 | 1.42070 |
| 0.2582 | 1.40419 | 0.9000 | 1.42246 |
| 0.3559 | 1.40647 | 0.9498 | 1.42417 |
| 0.4099 | 1.40779 | | |
| DMF (1) + DBA (2) ; $T$/K = 298.15 K | | | |
| 0.1038 | 1.41553 | 0.7709 | 1.42323 |
| 0.2076 | 1.41628 | 0.8398 | 1.42459 |
| 0.3608 | 1.41761 | 0.8989 | 1.42587 |
| 0.4897 | 1.41900 | 0.9497 | 1.42706 |
| 0.5933 | 1.42034 | 0.9738 | 1.42765 |
| 0.6880 | 1.42178 | | |
| DMF (1) + BA (2) ; $T$/K = 293.15 K | | | |
| 0.0599 | 1.40222 | 0.5059 | 1.41521 |
| 0.1056 | 1.40348 | 0.6043 | 1.41829 |
| 0.1591 | 1.40498 | 0.6569 | 1.41992 |
| 0.2010 | 1.40619 | 0.6999 | 1.42126 |
| 0.2505 | 1.40760 | 0.7537 | 1.42292 |
| 0.3017 | 1.40911 | 0.8036 | 1.42448 |
| 0.3583 | 1.41079 | 0.8579 | 1.42618 |
| 0.3992 | 1.41200 | 0.9048 | 1.42763 |
| 0.4401 | 1.41322 | 0.9546 | 1.42915 |
| DMF (1) + BA (2) ; *T*/K = 298.15 K | | | |
| 0.0575 | 1.39946 | 0.6005 | 1.41578 |
| 0.1092 | 1.40093 | 0.6608 | 1.41769 |
| 0.1519 | 1.40215 | 0.6934 | 1.41871 |
| 0.2043 | 1.40367 | 0.7572 | 1.42076 |
| 0.2448 | 1.40485 | 0.8056 | 1.42230 |
| 0.3090 | 1.40680 | 0.8546 | 1.42383 |
| 0.3558 | 1.40820 | 0.9076 | 1.42553 |
| 0.5028 | 1.41274 | 0.9553 | 1.42705 |
| DMF (1) + BA (2) ; *T*/K = 303.15 K | | | |
| 0.0505 | 1.39644 | 0.5281 | 1.41094 |
| 0.1477 | 1.39927 | 0.6056 | 1.41337 |
| 0.2019 | 1.40087 | 0.6574 | 1.41502 |
| 0.2410 | 1.40209 | 0.6960 | 1.41625 |
| 0.3024 | 1.40392 | 0.7558 | 1.41820 |
| 0.3558 | 1.40552 | 0.8006 | 1.41964 |
| 0.3992 | 1.40686 | 0.8544 | 1.42138 |
| 0.4358 | 1.40802 | 0.8998 | 1.42282 |
| 0.5114 | 1.41041 | 0.9466 | 1.42431 |
| DMF (1) + HxA (2) ; *T*/K = 298.15 K | | | |
| 0.0506 | 1.41613 | 0.5500 | 1.42107 |
| 0.1725 | 1.41707 | 0.6558 | 1.42257 |
| 0.2548 | 1.41779 | 0.7573 | 1.42411 |
| 0.3476 | 1.41870 | 0.8492 | 1.42564 |
| 0.4461 | 1.41980 | 0.9530 | 1.42747 |

[a] The standard uncertainties, $u$, are: $u(x_1) = 0.0008$; $u(T) = 0.03\,\text{K}$; $u(p) = 1\,\text{kPa}$. The relative combined expanded standard uncertainty (0.95 level of confidence), $U_{\text{rc}}$, is: $U_{\text{rc}}(n_{\text{D}}) =$ 0.0030.

Table 5

Coefficients $A_i$ and standard deviations, $\sigma\left(F^{E}\right)$ (eq. 9), for the representation of the $F^{E}$ property at temperature $T$ and pressure $p$ = 0.1 MPa for *N*,*N*-dimethylformamide (1) + amine (2) systems by eq. 8.

| System | $T$/K | Property [a] $F^{E}$ | $A_0$ | $A_1$ | $A_2$ | $A_3$ | $A_4$ | $\sigma\left(F^{E}\right)$ |
|---|---|---|---|---|---|---|---|---|
| DMF + DPA | 293.15 | $V_m^E$ | –1.121 | –0.23 | –0.37 | | | 0.005 |
| | | $n_D^E$ | 0.00540 | 0.0033 | 0.0013 | | | 0.00004 |
| | 298.15 | $V_m^E$ | –1.157 | –0.25 | –0.30 | | | 0.004 |
| | | $\kappa_S^E$ | –191.0 | –84.2 | –52.2 | –18 | | 0.11 |
| | | $c^E$ | 148.6 | 120.2 | 90 | 75 | 48 | 0.10 |
| | | $\alpha_p^E$ | –52.5 | | | | | 0.5 |
| | | $n_D^E$ | 0.0056 | 0.0028 | 0.0015 | | | 0.00004 |
| | 303.15 | $V_m^E$ | –1.192 | –0.21 | –0.37 | | | 0.005 |
| | | $n_D^E$ | 0.00583 | 0.0032 | 0.0020 | | | 0.00003 |
| DMF + DBA | 298.15 | $V_m^E$ | 0.071 | –0.304 | –0.30 | | | 0.0015 |
| | | $\kappa_S^E$ | –70.7 | –59.6 | –38 | –35 | –27 | 0.08 |
| | | $c^E$ | 62.2 | 65 | 53 | 74 | 61 | 0.15 |
| | | $n_D^E$ | 0.00020 | 0.00109 | 0.0010 | 0.0010 | 0.0008 | 0.000004 |
| DMF + BA | 293.15 | $V_m^E$ | –0.978 | –0.29 | –0.12 | | | 0.003 |
| | | $n_D^E$ | 0.00527 | 0.00229 | | | | 0.00002 |
| | 298.15 | $V_m^E$ | –1.052 | –0.29 | –0.24 | | | 0.003 |
| | | $\kappa_S^E$ | –167.6 | –36.0 | –17.0 | | | 0.09 |
| | | $c^E$ | 151.2 | 92 | 48 | | | 0.3 |
| | | $\alpha_p^E$ | –86 | 56 | –195 | –44 | | 0.6 |
| | | $n_D^E$ | 0.00532 | 0.00189 | | | | 0.00001 |
| | 303.15 | $V_m^E$ | –1.060 | –0.23 | –0.28 | | | 0.003 |
| | | $n_D^E$ | 0.00578 | 0.00197 | | | | 0.00002 |
| DMF + HxA | 298.15 | $V_m^E$ | –0.084 | –0.316 | –0.17 | –0.07 | | 0.0010 |
| | | $\kappa_S^E$ | –53.1 | –38.0 | –24.4 | –11 | | 0.05 |
| | | $c^E$ | 52.8 | 49 | 44 | 27 | | 0.12 |
| | | $n_D^E$ | 0.00020 | 0.00173 | 0.0013 | | | 0.000014 |

[a] $F^{E} = V_m^E$, units: $cm^3 \cdot mol^{-1}$; $F^{E} = c^E$, units: $m \cdot s^{-1}$; $F^{E} = \kappa_S^E$ units: $TPa^{-1}$; $F^{E} = \alpha_p^E$, units: $10^{-6} \cdot K^{-1}$.

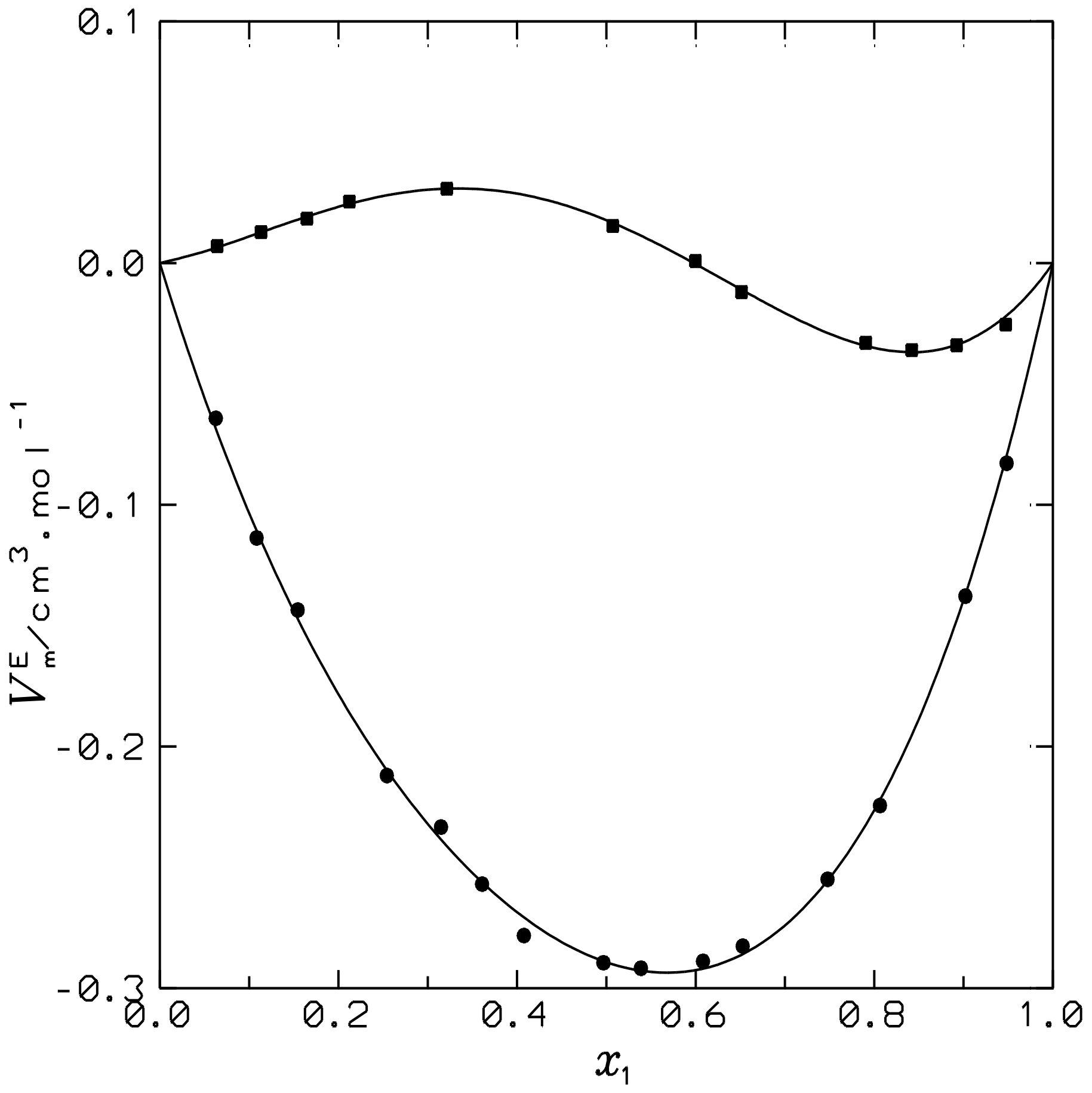

Figure 1

Excess molar volumes, $V_m^E$, for DMF (1) + DPA (2), or + DBA (2) systems at atmospheric pressure and 298.15 K. Full symbols, experimental values (this work): (⬢), DPA; (■), DBA. Solid lines, calculations with eq. 8 using the coefficients from Table 5.

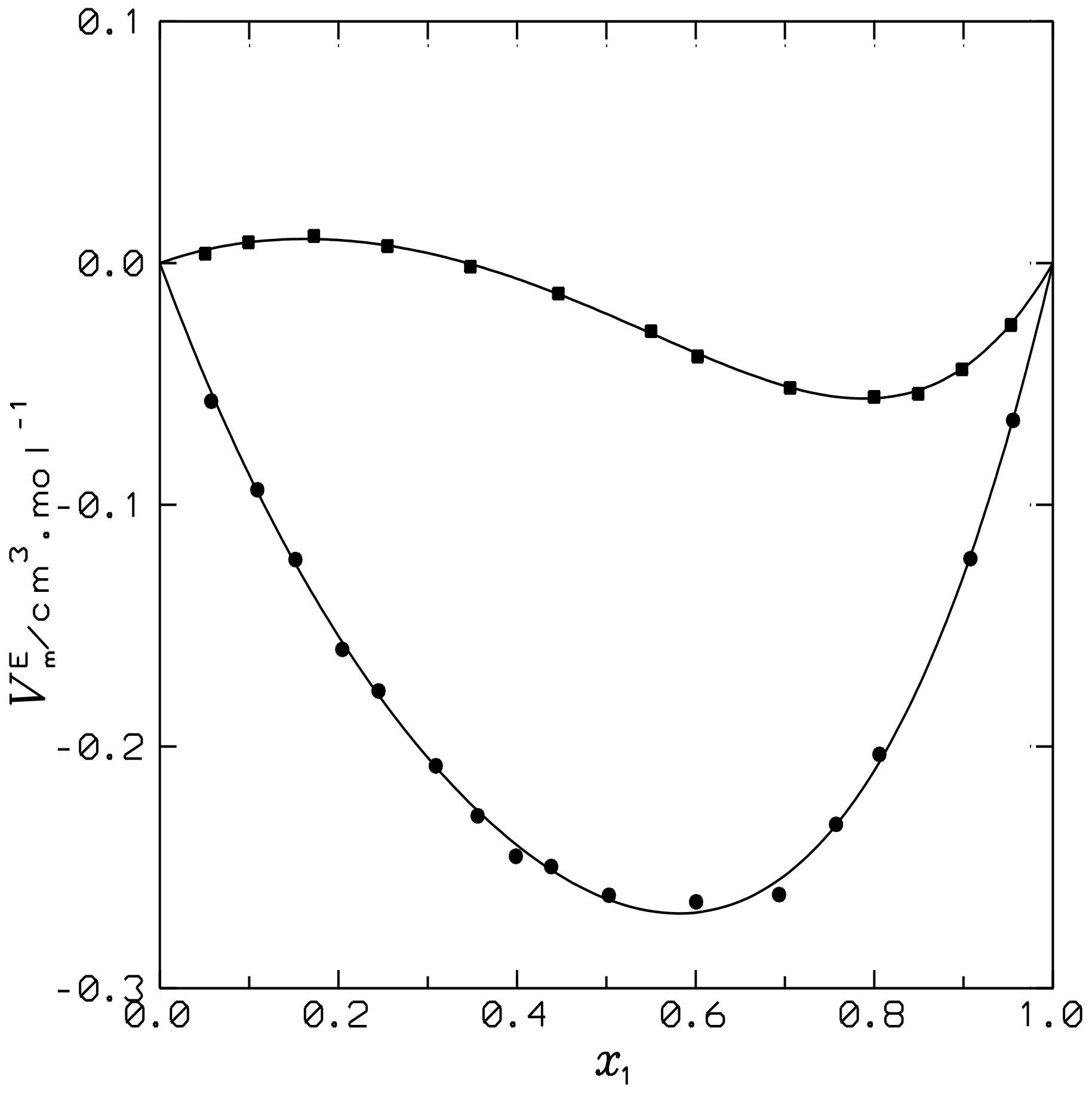

Figure 2

Excess molar volumes, $V_m^E$, for DMF (1) + BA (2), or + HxA (2) systems at atmospheric pressure and 298.15 K. Full symbols, experimental values (this work): (●), BA; (■), HxA. Solid lines, calculations with eq. 8 using the coefficients from Table 5.

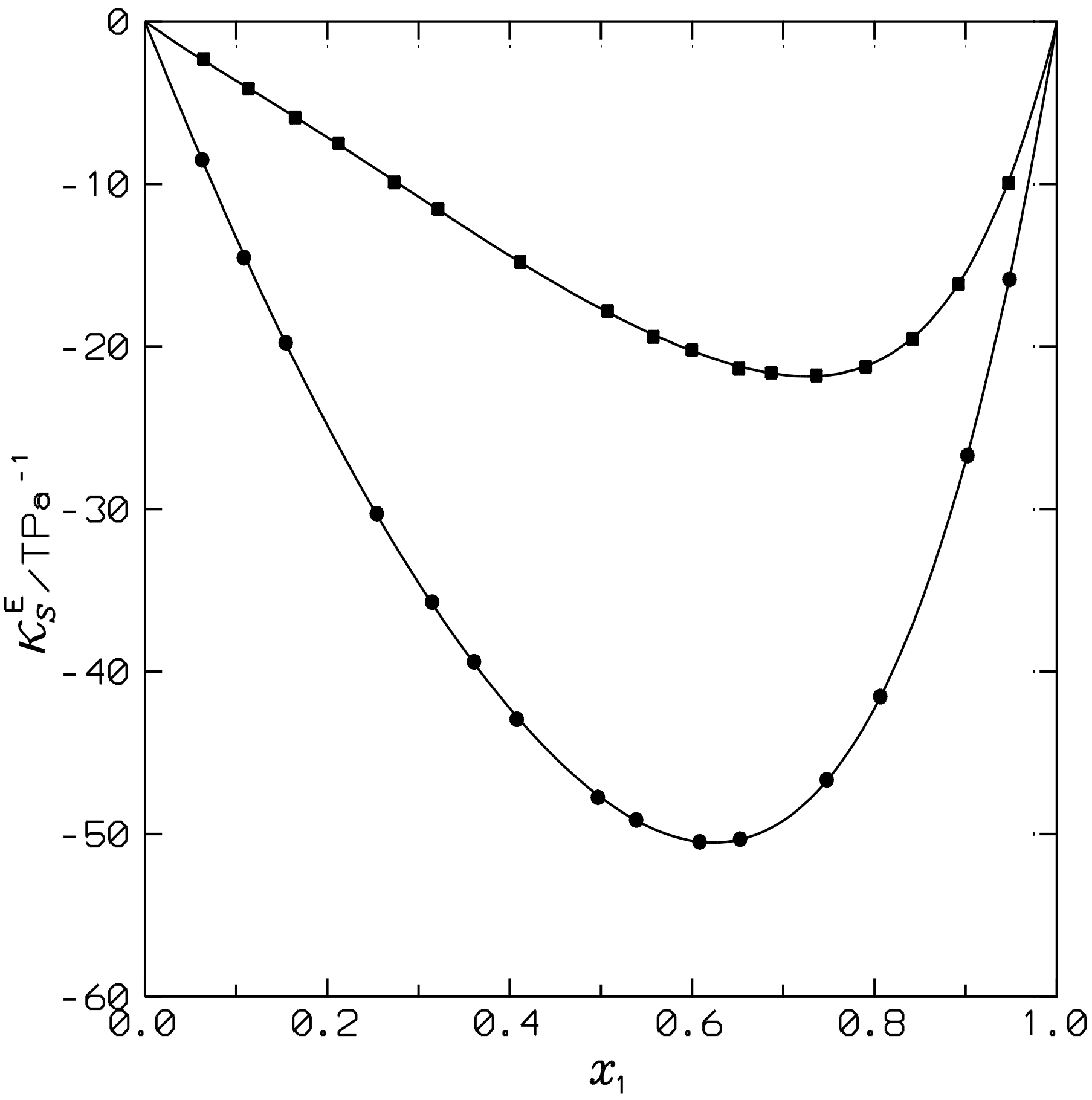

Figure 3

Excess isentropic compressibilities, $\kappa_S^{\mathrm{E}}$, for DMF (1) + DPA (2), or + DBA (2) systems at atmospheric pressure and 298.15 K. Full symbols, experimental values (this work): (●), DPA; (■), DBA. Solid lines, calculations with eq. 8 using the coefficients from Table 5.

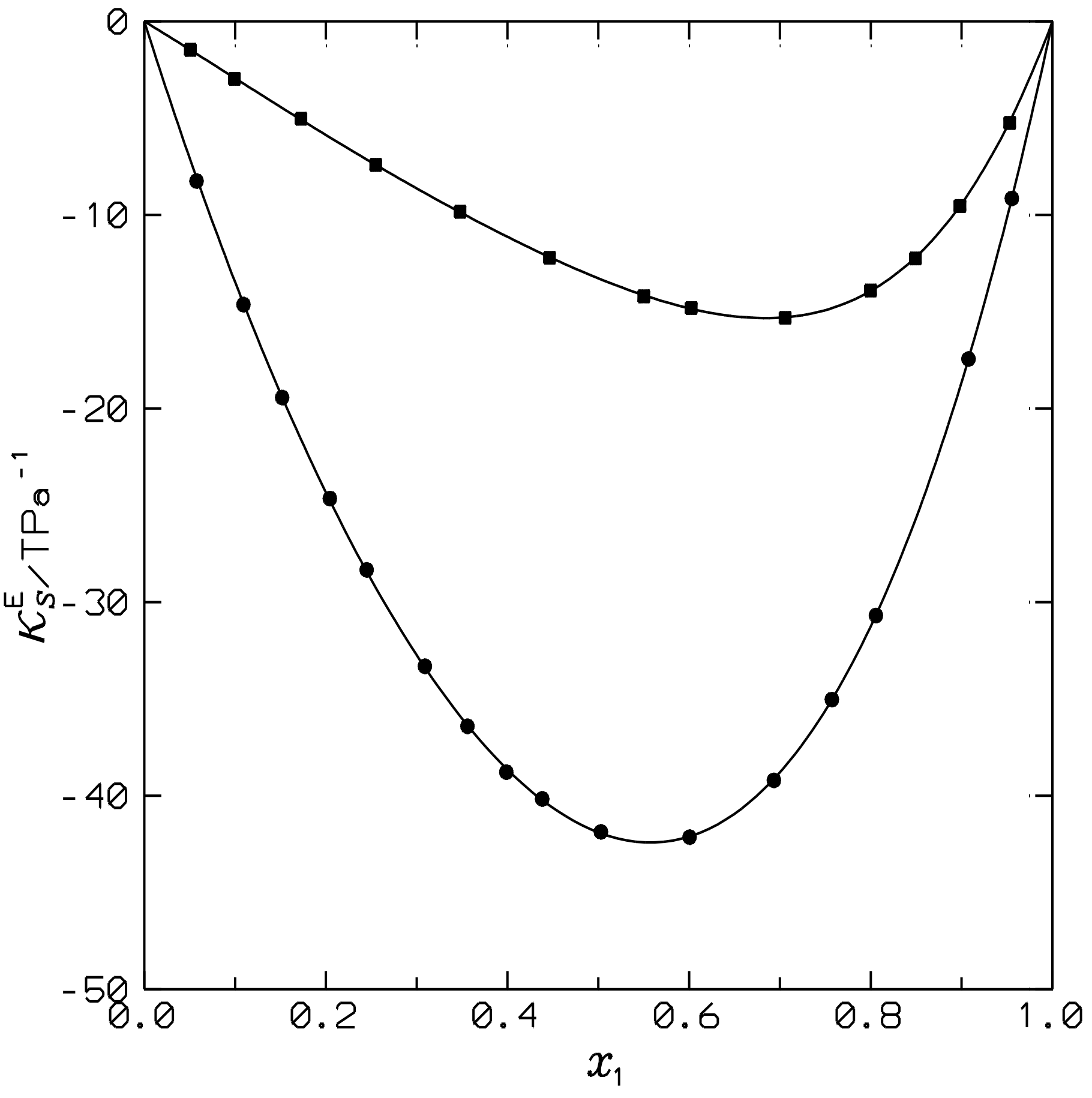

Figure 4

Excess isentropic compressibilities, $\kappa_S^{\mathrm{E}}$, for DMF (1) + BA (2), or + HxA (2) systems at atmospheric pressure and 298.15 K. Full symbols, experimental values (this work): (●), BA; (■), HxA. Solid lines, calculations with eq. 8 using the coefficients from Table 5.

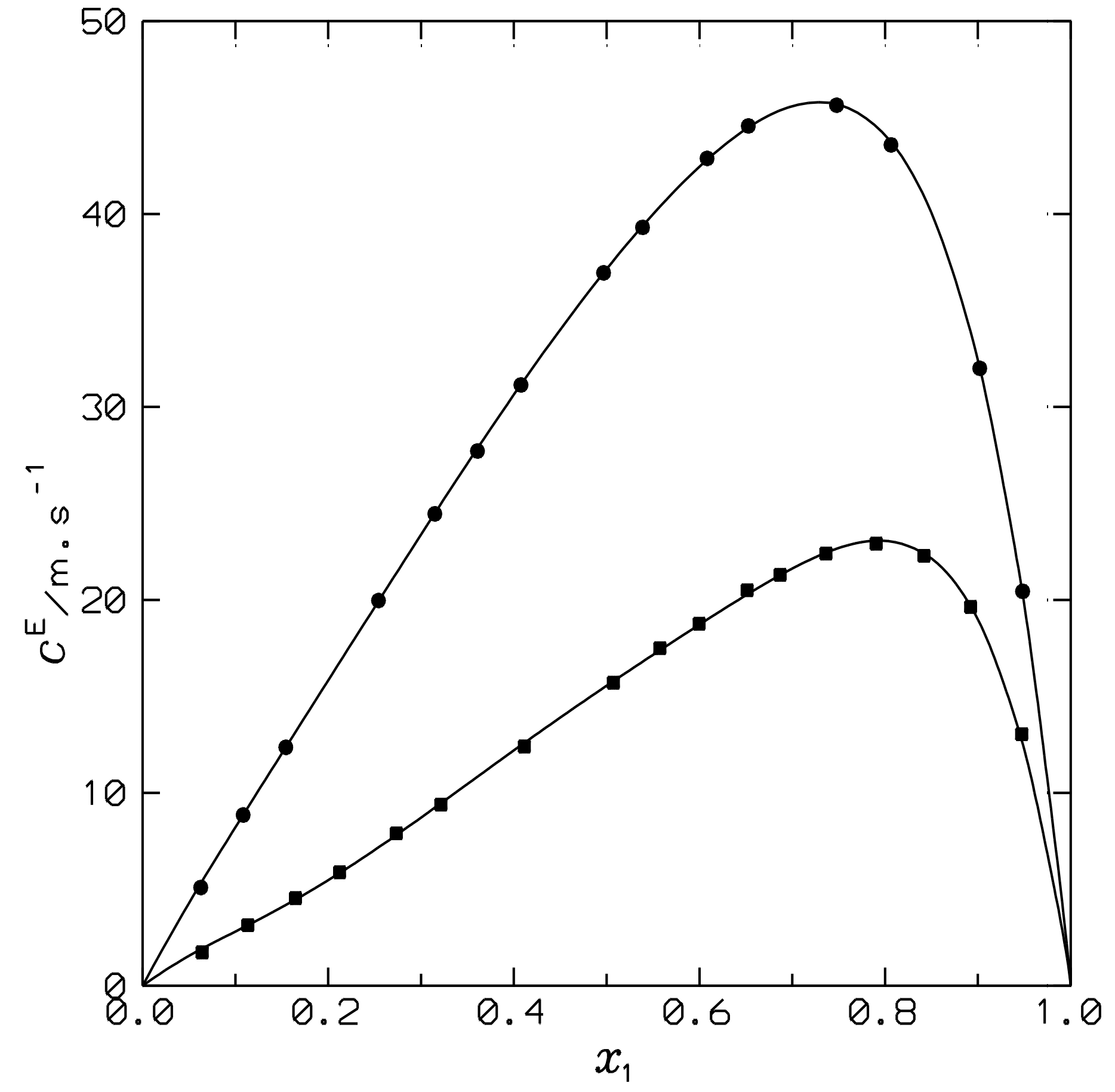

Figure 5

Excess speeds of sound, $c^{E}$, for DMF (1) + DPA (2), or + DBA (2) systems at atmospheric pressure and 298.15 K. Full symbols, experimental values (this work): (●), DPA; (■), DBA. Solid lines, calculations with eq. 8 using the coefficients from Table 5.

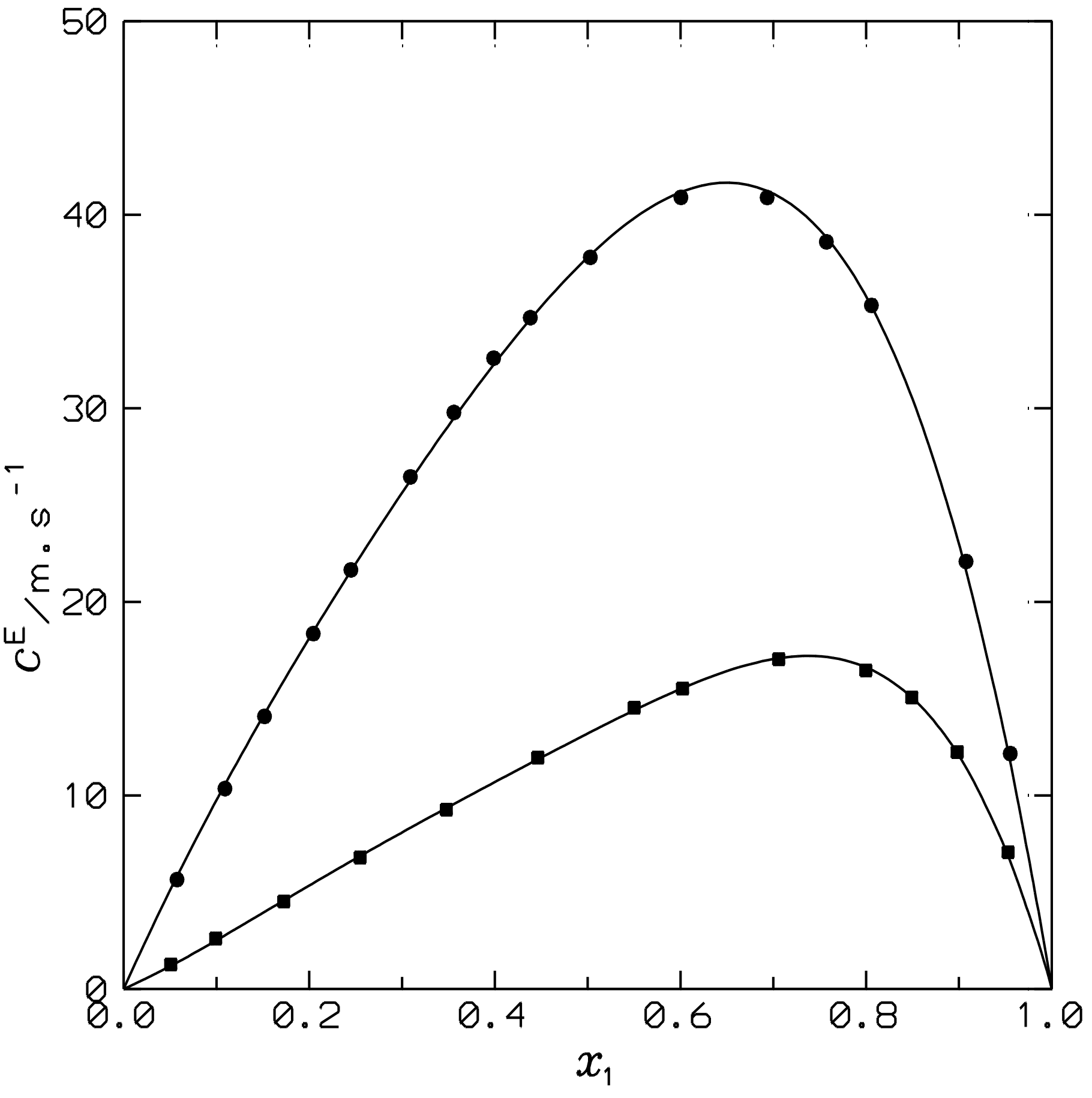

Figure 6

Excess speeds of sound, $c^E$, for DMF (1) + BA (2), or + HxA (2) systems at atmospheric pressure and 298.15 K. Full symbols, experimental values (this work): (●), BA; (■), HxA. Solid lines, calculations with eq. 8 using the coefficients from Table 5.

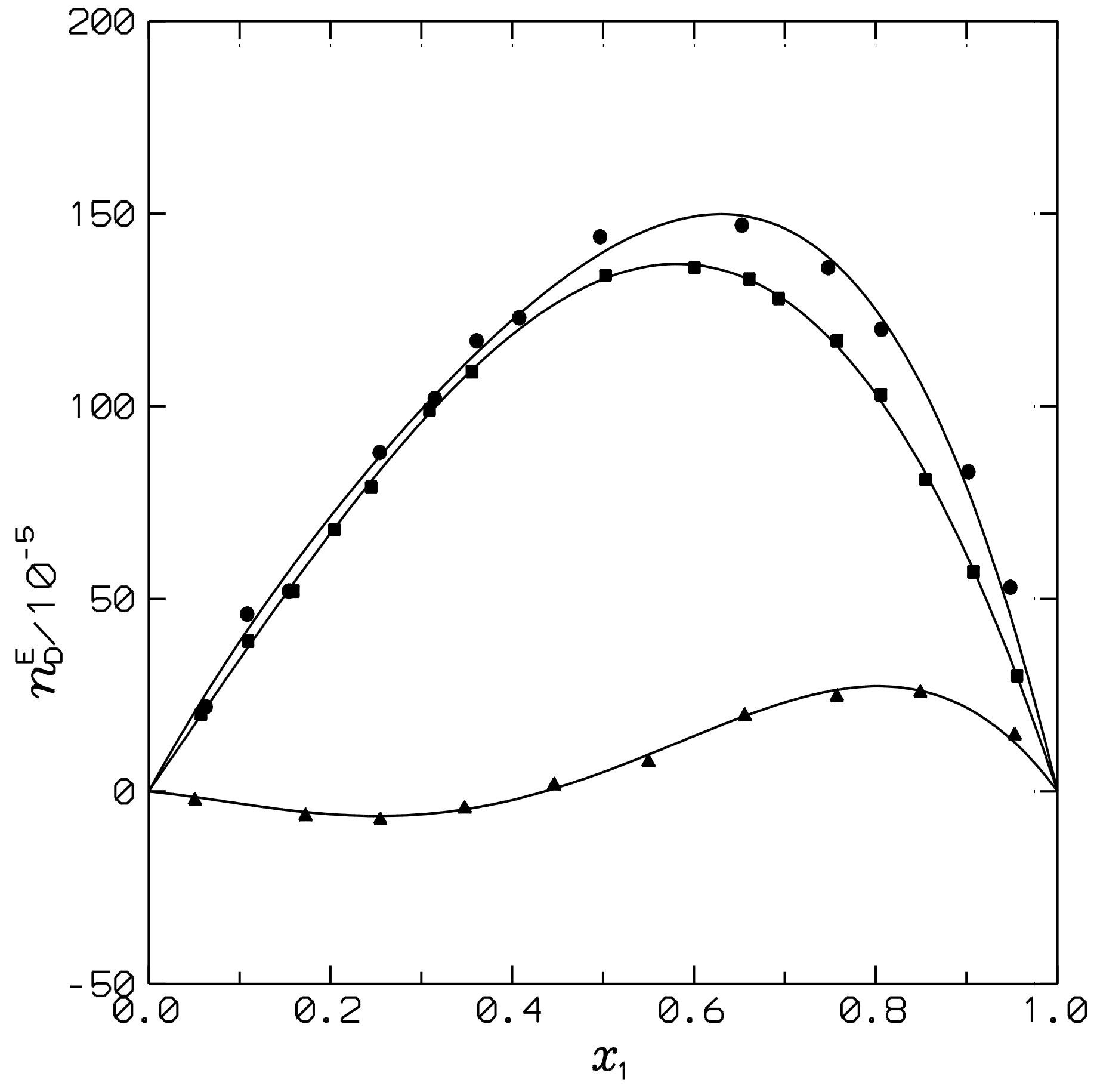

Figure 7

Excess refractive indices, $n_D^E$, for DMF (1) + amine (2) systems at atmospheric pressure and 298.15 K. Full symbols, experimental values (this work): (●), DPA; (■), BA; (▲), HxA. Solid lines, calculations with eq. 8 using the coefficients from Table 5.

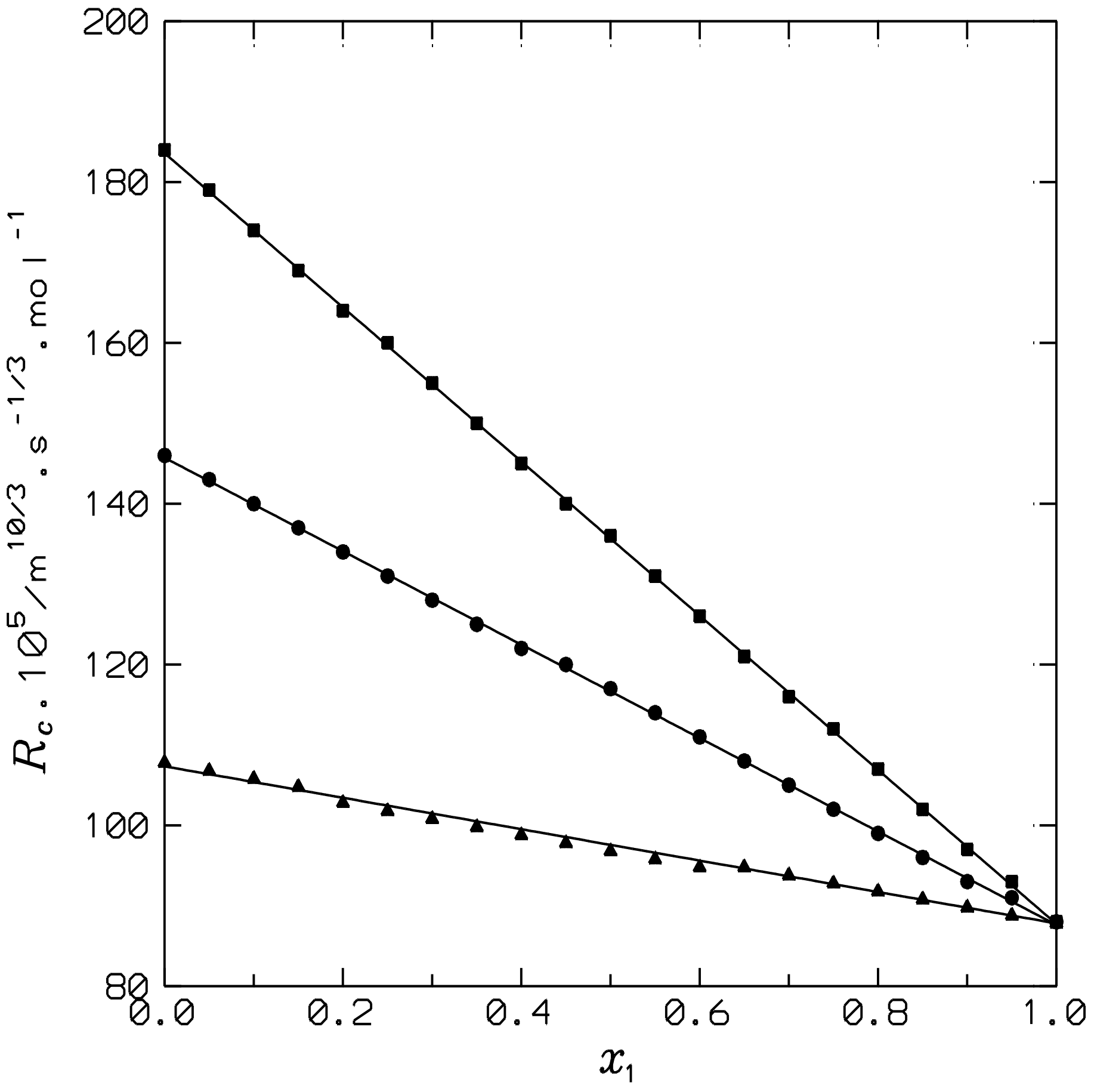

Figure 8

Rao's constant, $R_c$, for DMF (1) + amine (2) systems at atmospheric pressure and 298.15 K (this work): (●), DPA; (■), DBA; (▲), BA.

**FOR TOC ONLY**

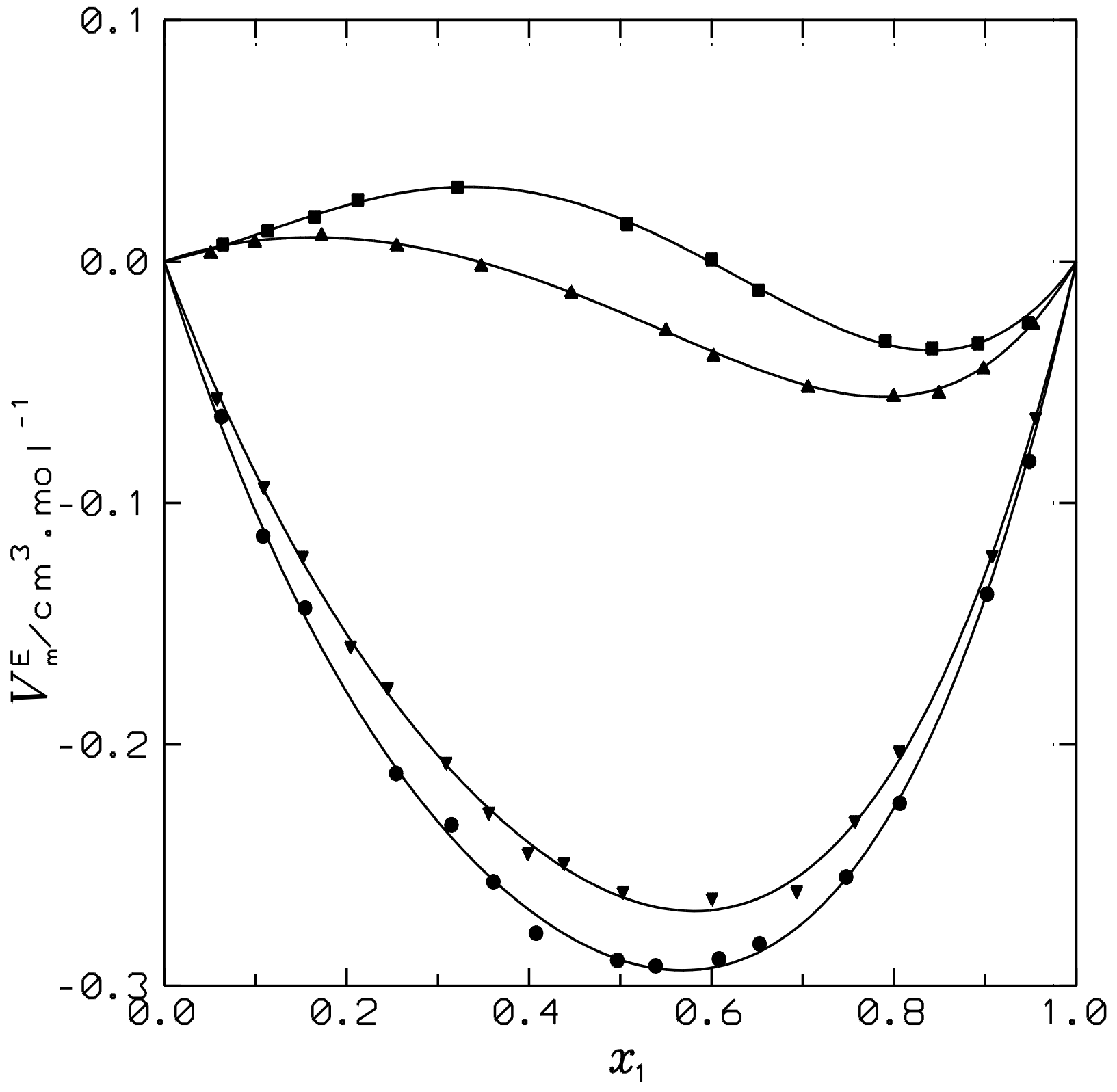

Excess molar volumes, $V_m^E$, for DMF (1) + amine (2) systems at atmospheric pressure and 298.15 K. Full symbols, experimental values: (●), DPA; (■), DBA; (▼), BA; (▲) HxA. Solid lines, results from the Redlich-Kister fittings.

## Supporting information

# Thermodynamics of amide + amine mixtures. 1. Volumetric, speed of sound and refractive index data for *N*,*N*-dimethylformamide + *N*-propylpropan-1-amine, + *N*-butylbutan-1-amine, + butan-1-amine, or + hexan-1-amine systems at several temperatures

Fernando Hevia, Ana Cobos, Juan Antonio González*, Isaías García de la Fuente, Luis Felipe Sanz

G.E.T.E.F., Departamento de Física Aplicada, Facultad de Ciencias, Universidad de Valladolid, Paseo de Belén, 7, 47011 Valladolid, Spain.

*e-mail: jagl@termo.uva.es; Fax: +34-983-423136; Tel: +34-983-423757

Table S1

Excess molar volumes, $V_m^E$, for *N*,*N*-dimethylformamide (1) + amine (2) mixtures at temperature *T* and pressure *p* = 0.1 MPa. [a]

| $x_1$ | $V_m^E$ /cm$^3$·mol$^{-1}$ | $x_1$ | $V_m^E$ /cm$^3$·mol$^{-1}$ |
|---|---|---|---|
| | DMF (1) + DPA (2) ; *T*/ K = 293.15 K | | |
| 0.0600 | − 0.0724 | 0.4974 | − 0.2847 |
| 0.1071 | − 0.1116 | 0.5487 | − 0.2869 |
| 0.1560 | − 0.1539 | 0.6562 | − 0.2787 |
| 0.1975 | − 0.1765 | 0.7514 | − 0.2413 |
| 0.2490 | − 0.1986 | 0.8216 | − 0.2017 |
| 0.3099 | − 0.2341 | 0.8501 | − 0.1875 |
| 0.3463 | − 0.2397 | 0.9025 | − 0.1447 |
| 0.3985 | − 0.2621 | 0.9486 | − 0.0800 |
| 0.4480 | − 0.2741 | | |
| | DMF (1) + DPA (2) ; *T*/ K = 298.15 K | | |
| 0.0626 | − 0.0642 | 0.5386 | − 0.2917 |
| 0.1083 | − 0.1137 | 0.6082 | − 0.2889 |
| 0.1544 | − 0.1435 | 0.6527 | − 0.2826 |
| 0.2541 | − 0.2120 | 0.7477 | − 0.2549 |
| 0.3148 | − 0.2333 | 0.8063 | − 0.2244 |
| 0.3609 | − 0.2569 | 0.9020 | − 0.1378 |
| 0.4077 | − 0.2782 | 0.9482 | − 0.0828 |
| 0.4966 | − 0.2895 | | |
| | DMF (1) + DPA (2) ; *T*/ K = 303.15 K | | |
| 0.0453 | − 0.0626 | 0.5415 | − 0.3027 |
| 0.1034 | − 0.1218 | 0.6016 | − 0.2975 |
| 0.1963 | − 0.1857 | 0.6517 | − 0.2908 |
| 0.2582 | − 0.2163 | 0.7502 | − 0.2558 |
| 0.3559 | − 0.2677 | 0.8498 | − 0.1931 |
| 0.4099 | − 0.2877 | 0.9000 | − 0.1465 |
| 0.4565 | − 0.2954 | 0.9498 | − 0.0838 |
| | DMF (1) + DBA (2) ; *T*/ K = 298.15 K | | |
| 0.0642 | 0.0070 | 0.5996 | 0.0009 |
| 0.1134 | 0.0128 | 0.6514 | − 0.0120 |
| 0.1647 | 0.0184 | 0.7904 | − 0.0330 |
| 0.2121 | 0.0255 | 0.8418 | − 0.0360 |
| 0.3213 | 0.0307 | 0.8921 | − 0.0340 |
| 0.5072 | 0.0153 | 0.9471 | − 0.0255 |
| DMF (1) + BA (2) ; *T*/ K = 293.15 K | | | |
| 0.0599 | − 0.0476 | 0.5483 | − 0.2483 |
| 0.1056 | − 0.0798 | 0.6569 | − 0.2449 |
| 0.1591 | − 0.1094 | 0.6999 | − 0.2320 |
| 0.2505 | − 0.1610 | 0.7537 | − 0.2128 |
| 0.3017 | − 0.1860 | 0.8036 | − 0.1858 |
| 0.3583 | − 0.2066 | 0.8579 | − 0.1587 |
| 0.3992 | − 0.2259 | 0.9048 | − 0.1102 |
| 0.5059 | − 0.2465 | 0.9546 | − 0.0587 |
| DMF (1) + BA (2) ; *T*/ K = 298.15 K | | | |
| 0.0575 | − 0.0571 | 0.4381 | − 0.2497 |
| 0.1092 | − 0.0938 | 0.5028 | − 0.2615 |
| 0.1519 | − 0.1227 | 0.6005 | − 0.2643 |
| 0.2043 | − 0.1598 | 0.6934 | − 0.2613 |
| 0.2448 | − 0.1770 | 0.7572 | − 0.2322 |
| 0.3090 | − 0.2080 | 0.8056 | − 0.2033 |
| 0.3558 | − 0.2287 | 0.9076 | − 0.1223 |
| 0.3986 | − 0.2454 | 0.9553 | − 0.0651 |
| DMF (1) + BA (2) ; *T*/ K = 303.15 K | | | |
| 0.0505 | − 0.0522 | 0.5281 | − 0.2698 |
| 0.1477 | − 0.1338 | 0.6056 | − 0.2704 |
| 0.2019 | − 0.1675 | 0.6960 | − 0.2574 |
| 0.2410 | − 0.1816 | 0.7558 | − 0.2289 |
| 0.3024 | − 0.2108 | 0.8006 | − 0.2107 |
| 0.3558 | − 0.2335 | 0.8544 | − 0.1664 |
| 0.4358 | − 0.2533 | 0.8998 | − 0.1251 |
| 0.5114 | − 0.2621 | 0.9466 | − 0.0748 |
| DMF (1) + HxA (2) ; *T*/ K = 298.15 K | | | |
| 0.0506 | 0.0039 | 0.6022 | − 0.0387 |
| 0.0992 | 0.0086 | 0.7056 | − 0.0516 |
| 0.1725 | 0.0112 | 0.7996 | − 0.0554 |
| 0.2548 | 0.0070 | 0.8492 | − 0.0541 |
| 0.3476 | − 0.0015 | 0.8982 | − 0.0439 |
| 0.4461 | − 0.0126 | 0.9530 | − 0.0256 |
| 0.5500 | − 0.0281 | | |

[a] The standard uncertainties are: $u(x_1) =$ 0.0008; $u(p) =$ 1 kPa; $u(T) =$ 0.02 K. The relative combined expanded standard uncertainty (0.95 level of confidence) is: $U_{rc}(V_m^E) = 0.025$.

Table S2

Isobaric thermal expansion coefficient, $\alpha_p$, and the corresponding excess function, $\alpha_p^{\mathrm{E}}$, at temperature $T$ = 298.15 K and pressure $p$ = 0.1 MPa, of *N*,*N*-dimethylformamide (1) + amine (2) mixtures. [a]

| $x_1$ | $\phi_1$ | $\alpha_p$ [b]/$10^{-3}$K$^{-1}$ | $\left(\frac{\partial\rho}{\partial T}\right)_p$ /kg m$^{-3}$ K$^{-1}$ | $r$ [c] | $\alpha_p^{\mathrm{E}}$ /$10^{-6}$K$^{-1}$ |
|---|---|---|---|---|---|
| | | DMF (1) + DPA(2) | | | |
| 0.0626 | 0.0361 | 1.228 | – 0.910953 | 0.999977 | – 4 |
| 0.1083 | 0.0638 | 1.220 | – 0.912081 | 0.999966 | – 5 |
| 0.1544 | 0.0930 | 1.211 | – 0.913352 | 0.999961 | – 7 |
| 0.2541 | 0.1605 | 1.192 | – 0.916600 | 0.999965 | – 11 |
| 0.3148 | 0.2050 | 1.181 | – 0.918932 | 0.999973 | – 11 |
| 0.3609 | 0.2407 | 1.172 | – 0.920886 | 0.999980 | – 12 |
| 0.4077 | 0.2787 | 1.162 | – 0.923030 | 0.999986 | – 13 |
| 0.4966 | 0.3564 | 1.144 | – 0.927515 | 0.999993 | – 13 |
| 0.5386 | 0.3959 | 1.135 | – 0.929800 | 0.999994 | – 13 |
| 0.6082 | 0.4656 | 1.120 | – 0.933767 | 0.999995 | – 12 |
| 0.6527 | 0.5134 | 1.109 | – 0.936379 | 0.999994 | – 12 |
| 0.7477 | 0.6245 | 1.086 | – 0.941942 | 0.999989 | – 9 |
| 0.8063 | 0.7003 | 1.070 | – 0.945181 | 0.999983 | – 8 |
| 0.9020 | 0.8378 | 1.042 | – 0.949609 | 0.999975 | – 4 |
| 0.9482 | 0.9113 | 1.027 | – 0.951097 | 0.999978 | – 2 |
| | | DMF (1) + BA(2) | | | |
| 0.0575 | 0.0452 | 1.283 | – 0.952253 | 1.000000 | – 14 |
| 0.1092 | 0.0868 | 1.262 | – 0.948543 | 0.999999 | – 23 |
| 0.1519 | 0.1219 | 1.248 | – 0.947139 | 0.999998 | – 26 |
| 0.2043 | 0.1660 | 1.232 | – 0.947004 | 0.999995 | – 29 |
| 0.2448 | 0.2008 | 1.221 | – 0.947790 | 0.999993 | – 29 |
| 0.3090 | 0.2574 | 1.205 | – 0.950026 | 0.999988 | – 28 |
| 0.3558 | 0.2998 | 1.193 | – 0.952015 | 0.999983 | – 27 |
| 0.3986 | 0.3394 | 1.183 | – 0.953824 | 0.999978 | – 25 |
| 0.4381 | 0.3767 | 1.173 | – 0.955318 | 0.999972 | – 24 |
| 0.5028 | 0.4394 | 1.156 | – 0.957074 | 0.999959 | – 22 |
| 0.6005 | 0.5382 | 1.128 | – 0.957395 | 0.999931 | – 20 |
| 0.6934 | 0.6368 | 1.098 | – 0.954824 | 0.999901 | – 20 |
| 0.7572 | 0.7074 | 1.076 | – 0.951809 | 0.999886 | – 21 |
| 0.8056 | 0.7626 | 1.059 | – 0.949272 | 0.999885 | – 21 |
| 0.9076 | 0.8839 | 1.027 | – 0.945455 | 0.999929 | – 16 |
| 0.9553 | 0.9431 | 1.014 | – 0.945764 | 0.999968 | – 11 |

[a] The standard uncertainties are: $u(x_1) =$ 0.0008; $u(p) =$ 1 kPa; $u(T) =$ 0.02 K. The relative combined expanded standard uncertainty (0.95 level of confidence) is $U_{rc}(\alpha_p^E) =$ 0.05.

[b] Density values at 293.15 and 303.15 K at the mole fractions reported at 298.15 K were obtained from the corresponding Redlich-Kister adjustments for $V_m^E$ .

[c] Regression coefficients (absolute values) obtained when fitting densities against temperature, assuming a linear dependence between the two quantities.

Table S3

Excess functions, at temperature $T$ = 298.15 K and pressure $p$ = 0.1 MPa, for $\kappa_S$, adiabatic compressibility, and $c$, speed of sound, of $N$,$N$-dimethylformamide (1) + amine (2) mixtures. [a]

| $x_1$ | $\kappa_S^E$ /TPa$^{-1}$ | $c^E$ /m·s$^{-1}$ | $x_1$ | $\kappa_S^E$ /TPa$^{-1}$ | $c^E$ /m·s$^{-1}$ |
|---|---|---|---|---|---|
| | | DMF (1) + DPA (2) | | | |
| 0.0626 | − 8.5 | 5.1 | 0.5386 | − 49.1 | 39.3 |
| 0.1083 | − 14.5 | 8.8 | 0.6082 | − 50.5 | 42.9 |
| 0.1544 | − 19.8 | 12.4 | 0.6527 | − 50.3 | 44.6 |
| 0.2541 | − 30.3 | 19.9 | 0.7477 | − 46.7 | 45.6 |
| 0.3148 | − 35.7 | 24.5 | 0.8063 | − 41.5 | 43.6 |
| 0.3609 | − 39.4 | 27.7 | 0.9020 | − 26.7 | 32.0 |
| 0.4077 | − 42.9 | 31.1 | 0.9482 | − 15.9 | 20.4 |
| 0.4966 | − 47.7 | 36.9 | | | |
| | | DMF (1) + DBA (2) | | | |
| 0.0642 | − 2.3 | 1.7 | 0.5574 | − 19.4 | 17.5 |
| 0.1134 | − 4.1 | 3.1 | 0.5996 | − 20.2 | 18.8 |
| 0.1647 | − 5.9 | 4.5 | 0.6514 | − 21.4 | 20.5 |
| 0.2121 | − 7.5 | 5.9 | 0.6869 | − 21.6 | 21.3 |
| 0.2734 | − 9.9 | 7.9 | 0.7361 | − 21.8 | 22.4 |
| 0.3213 | − 11.5 | 9.4 | 0.7904 | − 21.2 | 22.9 |
| 0.4114 | − 14.8 | 12.4 | 0.8418 | − 19.5 | 22.3 |
| 0.4526 | − 16.2 | 13.9 | 0.8921 | − 16.2 | 19.6 |
| 0.5072 | − 17.8 | 15.7 | 0.9471 | − 9.9 | 13.0 |
| | | DMF (1) + BA (2) | | | |
| 0.0575 | − 8.2 | 5.6 | 0.4381 | − 40.2 | 34.7 |
| 0.1092 | − 14.6 | 10.3 | 0.5028 | − 41.9 | 37.8 |
| 0.1519 | − 19.4 | 14.1 | 0.6005 | − 42.1 | 40.9 |
| 0.2043 | − 24.6 | 18.4 | 0.6934 | − 39.2 | 40.9 |
| 0.2448 | − 28.3 | 21.6 | 0.7572 | − 35.0 | 38.6 |
| 0.3090 | − 33.3 | 26.5 | 0.8056 | − 30.7 | 35.3 |
| 0.3558 | − 36.4 | 29.8 | 0.9076 | − 17.4 | 22.1 |
| 0.3986 | − 38.8 | 32.6 | 0.9553 | − 9.1 | 12.2 |
| | | DMF (1) + HxA (2) | | | |
| 0.0506 | − 1.5 | 1.3 | 0.6022 | − 14.8 | 15.5 |
| 0.0992 | − 2.9 | 2.6 | 0.7056 | − 15.3 | 17.0 |
| 0.1725 | − 5.0 | 4.5 | 0.7996 | − 13.9 | 16.5 |
| 0.2548 | − 7.4 | 6.8 | 0.8492 | − 12.2 | 15.1 |
| 0.3476 | − 9.8 | 9.3 | 0.8982 | − 9.5 | 12.2 |
| 0.4461 | − 12.2 | 11.9 | 0.9530 | − 5.2 | 7.1 |

| | | |
|---|---|---|
| 0.5500 | − 14.2 | 14.5 |

[a] The standard uncertainties, $u$, are: $u(x_1) = 0.0008$; $u(p) = 1$ kPa; $u(T) = 0.02$ K. The combined expanded standard uncertainties (0.95 level of confidence) are: $U_{rc}(c^E) =$ 0.0015; $U_{rc}(\kappa_S^E) =$ 0.05.

Table S4

Excess refractive indices, $n_D^E$, of *N*,*N*-dimethylformamide (1) + amine mixtures at temperature *T* and pressure *p* = 0.1 MPa. [a]

| $x_1$ | $n_D^E$ | $x_1$ | $n_D^E$ |
|---|---|---|---|
| DMF (1) + DPA (2) ; *T*/ K = 293.15 K | | | |
| 0.0600 | 0.00019 | 0.5487 | 0.00142 |
| 0.1556 | 0.00045 | 0.6018 | 0.00146 |
| 0.2614 | 0.00081 | 0.7033 | 0.00143 |
| 0.3463 | 0.00102 | 0.8216 | 0.00115 |
| 0.3985 | 0.00118 | 0.8844 | 0.00089 |
| 0.4554 | 0.00126 | 0.9486 | 0.00056 |
| DMF (1) + DPA (2) ; *T*/ K = 298.15 K | | | |
| 0.0626 | 0.00022 | 0.4966 | 0.00144 |
| 0.1083 | 0.00046 | 0.6527 | 0.00147 |
| 0.1544 | 0.00052 | 0.7477 | 0.00136 |
| 0.2541 | 0.00088 | 0.8063 | 0.00120 |
| 0.3148 | 0.00102 | 0.9020 | 0.00083 |
| 0.3609 | 0.00117 | 0.9482 | 0.00053 |
| 0.4077 | 0.00123 | | |
| DMF (1) + DPA (2) ; *T*/ K = 303.15 K | | | |
| 0.0453 | 0.00017 | 0.6016 | 0.00158 |
| 0.1034 | 0.00041 | 0.7502 | 0.00149 |
| 0.1963 | 0.00080 | 0.8498 | 0.00114 |
| 0.2582 | 0.00088 | 0.9000 | 0.00089 |
| 0.3559 | 0.00116 | 0.9498 | 0.00046 |
| 0.4099 | 0.00128 | | |
| DMF (1) + DBA (2) ; *T*/ K = 298.15 K | | | |
| 0.1038 | −0.00002 | 0.7709 | 0.00023 |
| 0.2076 | −0.00003 | 0.8398 | 0.00026 |
| 0.3608 | −0.00001 | 0.8989 | 0.00023 |
| 0.4897 | 0.00004 | 0.9497 | 0.00016 |
| 0.5933 | 0.00011 | 0.9738 | 0.00010 |
| 0.6880 | 0.00018 | | |
| DMF (1) + BA (2) ; *T*/ K = 293.15 K | | | |
| 0.0599 | 0.00019 | 0.5059 | 0.00128 |
| 0.1056 | 0.00034 | 0.6043 | 0.00137 |
| 0.1591 | 0.00051 | 0.6569 | 0.00135 |
| 0.2010 | 0.00066 | 0.6999 | 0.00131 |
| 0.2505 | 0.00078 | 0.7537 | 0.00119 |
| 0.3017 | 0.00093 | 0.8036 | 0.00106 |
| 0.3583 | 0.00107 | 0.8579 | 0.00087 |
| 0.3992 | 0.00114 | 0.9048 | 0.00064 |
| 0.4401 | 0.00120 | 0.9546 | 0.00033 |
| DMF (1) + BA (2) ; *T*/ K = 298.15 K | | | |
| 0.0575 | 0.00020 | 0.6005 | 0.00136 |
| 0.1092 | 0.00039 | 0.6608 | 0.00133 |
| 0.1519 | 0.00052 | 0.6934 | 0.00128 |
| 0.2043 | 0.00068 | 0.7572 | 0.00117 |
| 0.2448 | 0.00079 | 0.8056 | 0.00103 |
| 0.3090 | 0.00099 | 0.8546 | 0.00081 |
| 0.3558 | 0.00109 | 0.9076 | 0.00057 |
| 0.5028 | 0.00134 | 0.9553 | 0.00030 |
| DMF (1) + BA (2) ; *T*/ K = 303.15 K | | | |
| 0.0505 | 0.00020 | 0.5281 | 0.00147 |
| 0.1477 | 0.00057 | 0.6056 | 0.00146 |
| 0.2019 | 0.00075 | 0.6574 | 0.00142 |
| 0.2410 | 0.00092 | 0.6960 | 0.00136 |
| 0.3024 | 0.00107 | 0.7558 | 0.00127 |
| 0.3558 | 0.00117 | 0.8006 | 0.00113 |
| 0.3992 | 0.00126 | 0.8544 | 0.00093 |
| 0.4358 | 0.00134 | 0.8998 | 0.00069 |
| 0.5114 | 0.00146 | 0.9466 | 0.00040 |
| DMF (1) + HxA (2) ; *T*/ K = 298.15 K | | | |
| 0.0506 | −0.00002 | 0.5500 | 0.00008 |
| 0.1725 | −0.00006 | 0.6558 | 0.00020 |
| 0.2548 | −0.00007 | 0.7573 | 0.00025 |
| 0.3476 | −0.00004 | 0.8492 | 0.00026 |
| 0.4461 | 0.00002 | 0.9530 | 0.00015 |

[a] The standard uncertainties, *u*, are: $u(x_1) = 0.0008$; $u(T) = 0.03\,\mathrm{K}$; $u(p) = 1\,\mathrm{kPa}$. The relative combined expanded standard uncertainty (0.95 level of confidence), $U_{\mathrm{rc}}$, is: $U_{\mathrm{rc}}(n_{\mathrm{D}}^{\mathrm{E}}) = 0.04$.